\documentclass[prd,twocolumn,reprint,preprintnumbers,nofootinbib,superscriptaddress]{revtex4-1}
\input{header.tex}

\usepackage{xcolor}

\begin{document}
\reportnum{-2}{CERN-TH-2024-156}
\title{How new physics affects primordial neutrinos decoupling: Direct Simulation Monte Carlo approach}
\author{Maksym~Ovchynnikov}
\email{maksym.ovchynnikov@cern.ch}
\affiliation{Theoretical Physics Department, CERN, 1211 Geneva 23, Switzerland}
\affiliation{Institut für Astroteilchen Physik, Karlsruher Institut für Technologie (KIT), Hermann-von-Helmholtz-Platz 1, 76344 Eggenstein-Leopoldshafen, Germany}
\author{Vsevolod Syvolap}
\email{sivolapseva@gmail.com}
\affiliation{Instituut-Lorentz, Leiden University, Niels Bohrweg 2, 2333 CA Leiden, The Netherlands}

\date{\today}

\begin{abstract}
Cosmological observations from Big Bang Nucleosynthesis and the Cosmic Microwave Background (CMB) offer crucial insights into the Early Universe, enabling us to trace its evolution back to lifetimes as short as 0.01 seconds. Upcoming CMB spectrum measurements will achieve unprecedented precision, allowing for more accurate extraction of information about the primordial neutrinos. This provides an opportunity to test whether their properties align with the predictions of the standard cosmological model or indicate the presence of new physics that influenced the evolution of the MeV-temperature plasma. A key component in understanding how new physics may have affected primordial neutrinos is solving the neutrino Boltzmann equation. In this paper, we address this question by developing a novel approach -- neutrino Direct Simulation Monte Carlo (DSMC). We discuss it in-depth, highlighting its model independence, transparency, and computational efficiency -- features that current state-of-the-art methods lack. Then, we introduce a proof-of-concept implementation of the neutrino DSMC and apply it to several toy scenarios, showcasing key aspects of the primordial plasma's evolution in the presence of new physics.

\end{abstract}

\maketitle

\section{Introduction}
\label{sec:introduction}

Primordial neutrinos are an important messenger from the Early Universe, bringing us information about the state of the Universe at times as early as $t\ll 1\text{ s}$. Their direct detection is significantly more challenging than that of primordial photons due to their tiny interaction cross-section, which is governed by weak interactions. However, numerously populating the primordial plasma, they affected a number of cosmological observables. It makes it possible to indirectly extract information about their properties from precise cosmic measurements. In particular, they contribute to the number of ultrarelativistic (UR) degrees of freedom,
\begin{equation}
N_{\text{eff}} = \frac{8}{7}\left(\frac{11}{4}\right)^{\frac{4}{3}}\frac{\rho_{\text{UR}}-\rho_{\gamma}}{\rho_{\gamma}},
\label{eq:Neff}
\end{equation}
where $\rho_{\text{UR}}$ is the energy density of the ultrarelativistic species at the moment of CMB formation, and $\rho_{\gamma}$ is the energy density of photons. This quantity influences the Cosmic Microwave Background (CMB) and may be extracted from its measurements.

It is not only the total neutrino energy density that is important. Another essential property is the shape of the neutrino energy distribution function. It handles the neutron-to-proton conversion at MeV temperatures, which determines the onset of Big Bang Nucleosynthesis (BBN), as well as affects Baryon Acoustic Oscillations (BAO)~\cite{Bashinsky:2003tk,Baumann:2019keh}. The shape of the distribution may significantly modify the cosmological neutrino mass bound~\cite{Alvey:2021xmq}.

Assuming the standard cosmological history, based on the $\Lambda$CDM model, $N_{\text{eff}}$ is fully represented by neutrino, and its value is 3.043-3.044~\cite{Mangano:2001iu,Bennett:2019ewm,Bennett:2020zkv,Akita:2020szl,Froustey:2020mcq,Cielo:2023bqp,Drewes:2024wbw}. The shape of the neutrino distribution is very close to the Fermi-Dirac distribution, with tiny distortions in the high-energy tail. Finally, there is no asymmetry between neutrinos and antineutrinos. Altogether, it serves as an input to the Standard Big Bang Nucleosynthesis model, which predicts the helium abundance $Y_{p} = 0.247\pm 0.00017$ (see, e.g.,~\cite{Pisanti:2007hk,Pitrou:2018cgg}). These numbers agree with the current BBN and CMB observations. In particular, the measurements performed by the Planck collaboration~\cite{Planck:2018nkj} constrain $N_{\text{eff}} = 2.99^{+0.34}_{-0.33}$ at 95\% CL, whereas the primordial Helium abundance measurements are in a range 0.233-0.2573, obtained by combining the observations from the works~\cite{Aghanim:2018eyx,Izotov:2014fga,Aver:2015iza,Peimbert:2016bdg,Fernandez:2018xx,Valerdi:2019beb,Aver:2021rwi,Matsumoto:2022tlr}. 

However, the uncertainty window of these observations leaves room for sizeable deviations from standard neutrino properties that may potentially originate from the presence of new physics at temperatures $T_{\text{EM}}\lesssim 5\text{ MeV}$, when neutrinos start decoupling.\footnote{Here and below, $T_{\text{EM}}$ denotes the temperature of the electromagnetic plasma.} Examples of such scenarios include the presence of non-standard neutrino interactions~\cite{Archidiacono:2013dua,Forastieri:2015paa,deSalas:2021aeh,Du:2021idh}, a lepton asymmetry in the neutrino sector~\cite{Dolgov:2002ab,Grohs:2016cuu,EscuderoAbenza:2020cmq,Gelmini:2020ekg,Froustey:2021azz,Escudero:2022okz,Froustey:2024mgf}, a change in the expansion dynamics of the Universe, and the injection of non-thermal neutrinos by hypothetical Long-Lived Particles, or LLPs~\cite{Dolgov:2000jw,Hannestad:2004px,Ruchayskiy:2012si,Fradette:2017sdd,Fradette:2018hhl,Boyarsky:2020dzc,Sabti:2020yrt,Boyarsky:2021yoh,Mastrototaro:2021wzl,Rasmussen:2021kbf}. The accuracy of the CMB measurements will be significantly improved with the future observations with Simons Observatory~\cite{SimonsObservatory:2018koc} (which has started collecting the data on June 2024) and CMB-S4 mission~\cite{CMB-S4:2016ple}. They will be able to measure $N_{\text{eff}}$ with a percent precision, thus providing a unique potential to shed light on properties of the new physics or constrain it in case of the absence of deviations from $\Lambda$CDM.

Under certain approximations of neutrino oscillations, understanding the impact of the new physics effects on the neutrino properties requires solving the Boltzmann equation on the neutrino distribution function $f_{\nu_{\alpha}}$:
\begin{equation}
    \frac{\partial f_{\nu_{\alpha}}}{\partial t} - pH\frac{\partial f_{\nu_{\alpha}}}{\partial p} =  \mathcal{I}_{\text{coll},\alpha}[f_{\nu_{\alpha}},p]
   \label{eq:boltzmann}
\end{equation}
Here, $p$ is neutrinos' momentum, $H$ is the Hubble factor accounting for the expansion of the Universe, and $\mathcal{I}_{\text{coll},\alpha}$ is the collision integral that takes care of the microscopic of the thermalization.

The main approach considered in literature is to reduce the integration inside $\mathcal{I}_{\text{coll}}$ analytically as much as possible and convert the complex integrodifferential equation~\eqref{eq:boltzmann} into a system of the ordinary differential equations by discretizing the grid of the comoving momenta (see the pioneering work~\cite{Hannestad:1995rs} as well as later realizations~\cite{Grohs:2015tfy,Gariazzo:2019gyi,Akita:2020szl}, and references therein). The method has also been used to study some well-motivated scenarios with LLPs such as Heavy Neutral Leptons (HNLs)~\cite{Dolgov:2000jw,Ruchayskiy:2012si,Sabti:2020yrt,Boyarsky:2021yoh,Mastrototaro:2021wzl,Rasmussen:2021kbf} and particles in late reheating scenarios~\cite{Hannestad:2004px,Kanzaki:2007pd,Hasegawa:2019jsa}. 

However, several problems exist with this approach. First, it has a limited range of applicability, requiring analytic matrix elements for the processes and high reducibility of the dimensionality of the integration in $\mathcal{I}_{\text{coll},\alpha}$. Second, even within the case studies, its computational complexity quickly grows if high-energy neutrinos are present in the system. For instance, depending on the grid density, solving the Boltzmann equation under the presence of HNLs with masses just $\simeq 200\text{ MeV}$ (injecting neutrinos with energies up to 100 MeV) may take days~\cite{Sabti:2020yrt}.

In addition, the method itself is very complex. The analytic reduction of the collision integral is highly non-trivial, the comoving grid density has to be adjusted to the model's parameters, and solver stability must be carefully verified. An indirect consequence of this is that there is the existing discrepancy between the predictions of various neutrino Boltzmann codes for the behavior of $N_{\text{eff}}$ in the presence of the injection of high-energy neutrinos with energies well exceeding the plasma temperature. While some studies predict that injection of such neutrinos would increase $N_{\text{eff}}$~\cite{Dolgov:2000jw,Mastrototaro:2021wzl}, the others show the opposite~\cite{Boyarsky:2021yoh,Rasmussen:2021kbf,Ruchayskiy:2012si}).

In this paper, we address these issues by developing proof-of-principle of a novel approach to solving the neutrino Boltzmann equation based on the so-called Direct Simulation Monte Carlo (DSMC)~\cite{bird1978monte,Bird2003,ROOHI20161,stefanov2019basic}. Its basis is the numerical particle representation of the Boltzmann equation: one starts with a large number of particles obeying some initial condition in momentum and spatial spaces and then directly simulates their interactions to study the equilibration. Due to the straightforwardness of the method, DSMC directly calculates the linear functionals, e.g., the number and energy densities, velocities, etc., without any simplifications. The existing case studies describe the implementations of DSMC that efficiently simulate collisions of a number of particles as large as $10^{8}$~\cite{fonseca2008lect,gallis2014direct}. As we will see, the simplicity of the scheme describing the interactions and the absence of momentum binning automatically release the DSMC approach from most of the problems described above.

This work also serves as the companion to the paper~\cite{Ovchynnikov:2024xyd}, which presents a summary of the results.

The paper is organized as follows. In Sec.~\ref{sec:neutrinos}, we review the properties of the primordial plasma around the neutrino decoupling, considering both the standard cosmological scenario and setups with new physics. Sec.~\ref{sec:existing-approaches} is devoted to a discussion on the existing approaches to solve the neutrino Boltzmann equation. In Sec.~\ref{sec:DSMC-basics}, we describe the basics of the DSMC approach and, in particular, why it may be well-applicable to studying the dynamics of primordial neutrinos. Sec.~\ref{sec:DSMC-neutrinos} discusses the necessary modifications to the DSMC simulation required to study the primordial MeV plasma, and how they can be implemented. In Sec.~\ref{sec:current-implementation}, we present our proof-of-principle realization of the approach and different cross-checks we performed to validate it against well-defined scenarios. In Sec.~\ref{sec:applications}, we apply the developed approach to a few case studies simplifying various physics setups, highlighting the variety of the applicability of the neutrino DSMC and the importance of using full Boltzmann equations. Finally, in Sec.~\ref{sec:conclusions}, we make conclusions.

\section{Primordial plasma at MeV temperatures}
\label{sec:neutrinos}

As we discussed in the introduction, throughout this study, we mainly focus on the temperature domain $1\text{ MeV}\lesssim T_{\text{EM}}\lesssim 5\text{ MeV}$, where neutrinos are already partially decoupled at $T\simeq 5\text{ MeV}$~\cite{Sabti:2020yrt}, but still interact significantly with the EM particles and themselves. This necessitates a detailed understanding of the dynamics of these interactions.

At these temperatures, the primordial plasma consists of light particles -- neutrinos $\nu,\bar{\nu}$, electromagnetically (EM) interacting light particles (electrons $e^{-}$, positrons $e^{+}$, and photons $\gamma$), as well as baryons $B = p,n$. The thermal population of other particles, such as muons,  $\tau$ leptons, mesons, and excited baryon states, can be safely neglected, as they are too heavy to be abundantly present at this epoch.

The homogeneous and isotropic Universe expands with the rate $H(t) = \dot{a}(t)/a(t)$, where $a(t)$ is the scale factor, and $H$ is the Hubble parameter. Assuming spatial flatness and neglecting the dark energy contribution, we get
\begin{equation}
    H(t) = \frac{1}{M_{\text{Pl}}}\sqrt{\frac{8\pi}{3} \rho_{\text{Universe}}} 
    \label{eq:hubble}
\end{equation}
where $M_{\text{Pl}}$ is the Planck mass, and $\rho_{\text{Universe}}$ the total energy density of the Universe. 

To understand the scaling of $\rho_{\text{Universe}}$, we need to discuss different components of the primordial plasma and, in particular, their interactions. 

\subsection{EM plasma and nucleons}
Let us first consider the EM plasma. Examples of the processes are Compton scattering and electron-positron annihilation into a pair of photons. The corresponding rate well exceeds the Hubble parameter for times $t \lesssim 10^4\text{ s}$, which includes the period we are interested in.\footnote{The decoupling of EM particles happens much later. In particular, the EM particles' thermalization time is much shorter than any relevant timescale for the temperatures above $T_{\text{EM}}\gtrsim 1\text{ keV}$. At lower temperatures, for example, by injecting high-energy $e^{\pm},\gamma$s, we have a chance for them to photodisintegrate primordial nuclei before the EM particles thermalize~\cite{Kawasaki:2004qu}.} This means that the population of the EM particles can always be well described by just one quantity -- the temperature of the EM plasma $T_{\text{EM}}\equiv T$. 

The distribution function $f_{e^{\pm}}$ of electrons and positrons is Fermi-Dirac, while for photons it is Bose-Einstein:
\begin{align}
    f_{e^{\pm}}(p,T) =& \frac{1}{\exp\left[\frac{\sqrt{p^{2}+m_{e}^{2}}}{T}\right]+1}, \\ f_{\gamma}(p,T) =& \frac{1}{\exp\left[\frac{p}{T}\right]-1}, 
    \label{eq:f-FD}
\end{align}
with the electron's mass $m_{e}\approx 0.511\text{ MeV}$. The temperature $T_{\text{EM}}$ is related to the total energy density of the EM particles $\rho_{\text{EM}}$ by the formula
\begin{equation}
    \rho_{\text{EM}}(T_{\text{EM}}) =\rho_{e^{\pm}}(T_{\text{EM}}) + \rho_{\gamma}(T_{\text{EM}})
    \label{eq:rho-EM}
\end{equation}
Here, the energy densities of $e^{\pm},\gamma$ are
\begin{align}
    \rho_{e^{\pm}}(T_{\text{EM}}) =& g_{e^{\pm}}\int \frac{d^{3}\mathbf{p}}{(2\pi)^{3}} \sqrt{p^{2}+m_{e}^{2}} \ f_{e^{\pm}}(p,T_{\text{EM}}), \\ \rho_{\gamma}(T_{\text{EM}}) =& g_{\gamma}\int \frac{d^{3}\mathbf{p}}{(2\pi)^{3}} p f_{\gamma}(p,T_{\text{EM}}),
\end{align}
with the factors $g_{e^{\pm}} = 4$ and $g_{\gamma} = 2$ staying for the spin and charge degrees of freedom.

Due to the electroneutrality of the Universe, we may neglect the chemical potential of electrons at the temperatures of interest $T_{\text{EM}} \simeq 1\text{ MeV}$. Indeed, it is $\mu_{e^{\pm}}/T_{\text{EM}} \sim \eta_B \simeq 10^{-9}$, where $\eta_B$ is the baryon-to-photon ratio.\footnote{The asymmetry, obviously, becomes non-negligible after $e^{+}e^{-}$ annihilation, at $T_{\text{EM}}\lesssim m_{e}$, but then electrons become irrelevant for the dynamics of the primordial plasma.} 

In terms of $T_{\text{EM}}$, the Hubble factor~\eqref{eq:hubble} can be rewritten as
\begin{equation}
    H(T_{\text{EM}})\equiv \frac{T_{\text{EM}}^2}{M_{\text{pl}}^*}, \quad M_{\text{pl}}^* \approx \frac{M_{\text{pl}}}{1.66 \sqrt{g_{*}(T_{\text{EM}})}}
    \label{eq:Hubble-g}
\end{equation}
where $g_*$ the effective number of relativistic species: $g_{*} = \rho_{\text{Universe}}/\frac{\pi^{2}}{30}T_{\text{EM}}^{4}$, with $g_{i}$ being the number of spin and charge degrees of freedom. Assuming the $\Lambda$CDM scenario and that all the species are in perfect equilibrium, we have 
$g_* \approx g_{\gamma} + 7/8(g_{e} + g_{\nu}) = 10.75$.

Finally, the number density of baryons $B$ in the Early Universe, $n_{B}$, may be expressed in terms of the baryon-to-photon ratio $\eta_{B}$ and the photon number density: 
\begin{equation}
n_{B}(T_{\text{EM}}) =\eta_{B}(T_{\text{EM}})n_{\gamma}(T_{\text{EM}})
\end{equation}
Knowing the value of $\eta_{B}$ during the CMB formation, $\eta_{B,\text{Planck}} = 6.09\cdot 10^{-10}$~\cite{Planck:2018nkj}, and the dynamics of the Universe expansion, the temperature dependence of $\eta_{B}$ may be calculated as
\begin{equation}
    \eta_{B}(T_{\text{EM}}) = \eta_{B,\text{Planck}} \times \left(\frac{a(T_{\text{EM,CMB}})T_{\text{EM,CMB}}}{a(T_{\text{EM}})T_{\text{EM}}}\right)^{3},
    \label{eq:eta-B}
\end{equation}
where $a$ is the scale factor of the Universe. The scaling of $\eta_{B}$ comes from the behavior of the baryon number density, $n_{B}\propto a^{-3}$, and the number density of photons, $n_{\gamma}\propto T_{\text{EM}}^{3}$. The resulting temperature-dependent factor stays for the entropy dilution of the Universe. In the standard cosmological scenario, Eq.~\eqref{eq:eta-B} gives $\eta_{B}(T_{\text{EM}}\simeq 1\text{ MeV})\approx 1.67\cdot 10^{-9}$.

The relative ratio between protons and neutrons, important for BBN, is handled by their weak interactions with neutrinos and $e^{\pm}$ particles, which drive the $p\leftrightarrow n$ conversion, so the baryons are coupled to the UR content of the plasma. However, because of the tiny number density and the absence of other hadrons in the plasma in the standard scenario, nucleons play a negligible role in the thermodynamics of the Universe at MeV temperatures.

\subsection{Neutrinos}
Let us now discuss neutrinos. Generically, there may be an asymmetry between neutrinos and antineutrinos, but the minimal cosmological setup assumes zero asymmetry.\footnote{Or at the level of baryon-to-photon ratio, which is negligible.} Neutrinos interact with themselves and $e^{\pm}$ particles via the weak force. The dimensional estimate for the weak interaction rates gives 
\begin{equation}
\Gamma_{\text{weak}} \simeq n_{\nu}\cdot \langle\sigma v\rangle \sim G_{F}^2 T_{\text{EM}}^5,
\end{equation}
where we assumed that the neutrinos have the thermal equilibrium with the EM plasma and, for the moment, set the neutrino temperature $T_{\nu}= T_{\text{EM}}$. Namely, $n_{\nu}\propto T_{\text{EM}}^{3}$ is the neutrino number density, while $\langle\sigma v\rangle$ is the thermally averaged cross-section, which scales as 
\begin{equation}
\langle\sigma v\rangle \sim G_{F}^{2}\langle s\rangle \sim G_{F}^{2}T_{\text{EM}}^{2},
\end{equation}
and $G_{F} \approx 1.167\cdot 10^{-5}\text{ GeV}^{-2}$ is the Fermi coupling. The important feature is that the cross-section scales with the energies of the interacting particles (we will return to it in Sec.~\ref{sec:applications}).

The rate becomes comparable to the Hubble expansion rate of the Universe already at $T_{\text{EM}}\sim 1\text{ MeV}$. As a result, at these temperatures, the weak reactions are no longer able to maintain equilibrium in the neutrinos sector, and the latter gradually decouple~\cite{Dolgov:2002wy}. The shape of their spectrum in $\Lambda$CDM closely follows the Fermi-Dirac one. Its temperature $T_{\nu_{\alpha}}$ remains equal to the EM temperature until the annihilation of electron-positron pairs, which happens around $T_{\text{EM}} \simeq m_{e}$. Then, their relation may be found from the entropy conservation law, giving $T_{\nu_{\alpha}} \approx (4/11)^{1/3}T_{\text{EM}}$. Extended neutrino decoupling introduces a small correction to this relation, leading to the value of $N_{\text{eff}}$ that is slightly larger than the instant decoupling result $N_{\text{eff}} = 3$.

\textbf{Neutrino interaction processes.} Let us now discuss neutrino interactions in more detail. They include elastic scatterings off neutrinos and $e^{\pm}$ and annihilations:
\begin{align}
        \quad \nu_\alpha + e^{\pm} \leftrightarrow& \nu_\alpha + e^{\pm}, \quad
         \nu_\alpha + \bar{\nu}_\alpha \leftrightarrow e^- + e^+, \\  \quad \nu_\alpha + \nu_\beta \leftrightarrow& \nu_\alpha + \nu_\beta, \quad
        \nu_\alpha + \bar\nu_\alpha \leftrightarrow \nu_\beta + \bar{\nu}_\beta,
        \label{eq:neutrino-interactions}
\end{align}
as well as charge-conjugated ones~\cite{Sabti:2020yrt}. The other reactions include the electroweak corrections, such as sub-dominant $e^{+}e^{-}\to \nu_{\alpha}\bar{\nu}_{\alpha}\gamma$.

The MeV plasma is ``flavor-asymmetric'' in the sense that electrons and positrons are present in plasma, while $\mu$ and $\tau$ leptons are not. Given the structure of the charged current of weak interactions, which includes the lepton and the corresponding neutrino, the direct interaction rate of $\nu_{e}$ with $e^{\pm}$ is larger than the rate of the corresponding scatterings but with $\nu_{\mu,\tau}$. Because of this, one can naively expect that $\nu_{\mu,\tau}$s decouple earlier from the EM plasma, while $\nu_{e}$s are kept longer in equilibrium. However, besides the interactions~\eqref{eq:neutrino-interactions}, neutrinos also experience flavor transitions called oscillations. The oscillations generically appear because the neutrino charge eigenstates do not coincide with the mass eigenstates. 

In the primordial plasma, the neutrino oscillation rate is severely affected by the dense medium. Namely, neutrinos acquire a correction to the self-energy caused by interactions with electrons and positrons~\cite{Dolgov:2002wy}. It effectively translates to a potential $\mathcal{V}^{(\nu_{\alpha})}_{\text{eff}}$ in the Hamiltonian describing the propagation of neutrinos $\nu_{\alpha}$. The functional form of the potential is $\mathcal{V}^{(\nu_{\alpha})}_{\text{eff}} = C^\alpha \frac{G_F^2 T_{\text{EM}}^4 E_{\nu}}{\alpha_{\text{EM}}}$, where $C^\alpha$ is neutrino-dependent constant. 

If $\mathcal{V}^{(\nu_{\alpha})}_{\text{eff}}$ is higher than the energy splitting for different neutrino eigenstates $\Delta m^2/2E_{\nu}$, the mixing angle is effectively suppressed, and oscillations can be ignored.  Therefore, the oscillations are absent at high temperatures and/or for high-energy neutrinos. In $\Lambda$CDM, they effectively turn on at $T_{\text{EM}}\simeq 3\text{ MeV}$. 

In total, because of oscillations, the interactions of three neutrino flavors with the EM plasma are similar. Therefore, the decoupling of $\nu_{e},\nu_{\mu},\nu_{\tau}$ occurs in a similar fashion.

\subsection{How new physics may spoil properties of primordial plasma}

There are various ways of introducing new physics to the primordial plasma. They will change the dynamics of the primordial plasma, in particular, departing the neutrino properties from the $\Lambda$CDM ones. To be specific, let us consider the scenario appearing in many well-motivated extensions of the Standard Model, adding LLPs (with mass $m\gg T_{\text{EM}}$). 

To significantly affect the Universe, such particles need to be out-of-equilibrium relics. Before decaying, they would increase the energy density of the Universe and hence modify the Hubble factor. After decaying, their influence gets split into many contributions. First, they still modify the dynamics of the Universe by introducing the dilution to the scale factor $a(T)$. It influences the behavior and value of $\eta_{B}$ at MeV temperatures via Eq.~\eqref{eq:eta-B}.\footnote{In particular, the presence of additional energy at high temperatures compared to $\Lambda$CDM leads to an increase of the scale factor at the CMB epoch, $a(T_{\text{CMB}})$. As $\eta_{B}(T_{\text{CMB}})$ is fixed, $\eta_{B}(T\simeq 1\text{ MeV})$ from Eq.~\eqref{eq:eta-B} must be larger than the $\Lambda$CDM value to compensate for the dilution.}

Second, their decay products may either constitute additional species (``dark radiation'') or inject energy into the population of neutrinos and the EM particles. The EM population gets immediately thermalized, which results in an increase of $T_{\text{EM}}$, while the neutrino injections cause the spectral distortions. Since neutrinos with different energies interact at different rates, much slower than the EM particles, the distortions will not disappear, affecting the total neutrino number and energy densities, as well as the $p\leftrightarrow n$ conversion rates.

Under such scenarios, the nucleons may also be involved in the thermodynamics of the Universe in a non-trivial way. Decaying LLPs may inject relatively long-lived mesons such as $\pi^{\pm}, K^{\pm}, K_{L}$. Before decaying, these particles experience numerous interactions with the SM plasma particles and themselves. Scattering off nucleons surprisingly becomes very efficient -- the smallness of $\eta_{B}$ is compensated by the largeness of the nucleon interaction cross-section, driven by the strong force~\cite{Boyarsky:2020dzc}. Because of these scatterings, the mesons change the distribution of their energy among the neutrino and EM sector~\cite{Akita:2024ork}, which leads to the impact on the time-temperature relation $t(T_{\text{EM}})$ and neutrino properties.  

\section{Existing approaches to solve the $\nu$ Boltzmann equation}
\label{sec:existing-approaches}

In general, to study the thermalization of neutrinos, one has to solve the quantum kinetic equations (QKEs) for the neutrino density matrix~\cite{McKellar:1992ja,Sigl:1993ctk,Vlasenko:2013fja,Gariazzo:2019gyi,Froustey:2020mcq,Akita:2020szl}. However, for our purposes, it may be reasonable to approximate the oscillations by the temperature-dependent oscillation probabilities, $\langle P_{\alpha\beta}\rangle(E_{\nu},T_{\text{EM}})$, similarly to how this is done in~\cite{Ruchayskiy:2012si,Sabti:2020yrt}. Then, it may be possible to reduce the complexity by converting the QKEs into the Boltzmann equations for the neutrino distribution function $f_{\nu_{\alpha}}$ in the momentum space:
\begin{multline}
\frac{\partial f_{\nu_{\alpha}}(E_{\nu},t)}{\partial t} - E_{\nu} H\frac{\partial f_{\nu_{\alpha}}(E_{\nu},t)}{\partial E_{\nu}} = \\ =\sum_{\beta}\langle P_{\beta\alpha}\rangle\cdot \mathcal{I}_{\text{coll},\nu_{\beta}}[E_{\nu},f_{\nu_{\alpha}},f_{\nu_{\beta}},T],
\label{eq:boltzmann-equation}
\end{multline}
supplemented with the Friedmann equation describing the expansion of the Universe (and in particular $H$), the equation for the evolution of the EM plasma temperature $T_{\text{EM}}$, and the equation governing the dynamics of LLPs in case they are present. Here, $E_{\nu_{\alpha}} 
= |\mathbf{p}_{\nu}|$ is the neutrino physical momentum. $\mathcal{I}_{\text{coll},\nu_{\beta}}$ is the collision integral for the neutrino of the flavor $\beta$, which in general contains a source term from new physics particles, a neutrino-neutrino interaction term, and a neutrino-EM interaction term. It has the form~\cite{Grohs:2015tfy}
\begin{multline}
\mathcal{I}_{\text{coll},\nu_{\alpha}} = \frac{1}{2E_{\nu_\alpha}}\sum \int \prod_{i=2} \frac{d^3 \mathbf{p}_i}{(2\pi)^3 2E_i} \prod_{f=1} \frac{d^3 \mathbf{p}_f}{(2\pi)^3 2E_f} \\ 
\times |\mathcal{M}|^2F[f](2\pi)^4\delta^{(4)} \left(\sum_{i=1} p_i- \sum_{f=1} p_f \right).
\label{Collterm}
\end{multline}
The first summation encompasses all potential interaction processes involving $\nu_\alpha$, with $i = 1$ representing the neutrino itself ($p_{1} \equiv (E_{\nu_{\alpha}},\mathbf{p}_{1})$). The integral extends over all possible states of $\nu_\alpha$ characterized by momentum $p_1$. Here, $i$ and $j$ denote the initial and final states of a given process, respectively. The term $|\mathcal{M}|^2$ represents the squared matrix element of the process (see Table 3 in ref.~\cite{Sabti:2020yrt} for the explicit expressions of $|\mathcal{M}|^2$ relevant to neutrino processes at MeV-scale temperatures). The factor $F[f]$ accounts for the statistical distribution within the medium and is given by
\begin{equation}
F[f] = \prod_{i=1}(1 \mp f_i)\prod_{f=1} f_f - \prod_{i=1} f_i \prod_{f=1} (1 \mp f_f),
\end{equation}
where $f_{i,f}$ denote the momentum distributions for the $i$-th and $f$-th particles. Finally, the factor $(1 - f)$ corresponds to Pauli blocking for fermions, whereas $(1 + f)$ corresponds to Bose enhancement for bosons. Finally, the $\delta$ function ensures the conservation of the 4-momentum in the process.

Depending on the scenario studied, there are two different state-of-the-art approaches to solving the Boltzmann equation~\eqref{eq:boltzmann-equation}. If the neutrinos injected by decays of new physics are close to thermal, $E_{\nu}\simeq 3.15T_{\text{EM}}$, or if decays are solely electromagnetic, then it may be possible to approximate the neutrino distribution by
\begin{equation}
    f_{\nu_{\alpha}}(E_{\nu},t) \approx f_{\text{FD}}(E_{\nu},T_{\nu_{\alpha}}(t)) = \frac{1}{\exp\left[\frac{E_{\nu}}{T_{\nu_{\alpha}}(t)}\right]+1}
    \label{eq:distribution-thermal}
\end{equation}
and consider an integrated version of the Boltzmann equations on the three neutrino temperatures $T_{\nu_{\alpha}}(t)$~\cite{Escudero:2018mvt,EscuderoAbenza:2020cmq}. In the limit of negligible electron mass, it may be possible to represent the energy transfer rates in the system as an analytic expression. Another example includes obtaining a correction to neutrino high-energy tail caused by non-instant decoupling in the standard scenarios~\cite{Dolgov:1997mb}.\footnote{It may be converted to the momentum-dependent correction to the neutrino temperature $T_{\nu}(p)$ that approaches $T_{\text{EM}}$ at $p\gg 3.15 \cdot T_{\text{EM}}$ and vanishes at small momenta.}

In practice, once we add new physics, the assumption of the perfect thermality of the neutrino distribution is typically violated. The first reason is the energy dependence of the equilibration of neutrinos, as discussed in Sec.~\ref{sec:neutrinos}. Neutrinos with different energies interact at very different rates, which leads to neutrino spectral distortions even if we simply heat the EM plasma. 

To study the distortions, one needs to solve the Boltzmann equation~\eqref{eq:boltzmann-equation} in the full generality. In the literature, this is done using the approach that we will call the discretization method. The algorithm is to analytically reduce the dimensionality of the integration in $\mathcal{I}_{\text{coll}}$ to some integer $k$ and then discretize the comoving momentum space $y = p\cdot a(t)$.\footnote{In the Standard cosmological scenario case, the comoving grid is convenient since it ``freezes'' the neutrino distribution: e.g., the peak of the energy distribution corresponds to the same $y$ at different times.} The integrodifferential Boltzmann equation is then converted into a system of ordinary differential equations (see~\cite{Hannestad:1995rs,Grohs:2015tfy,Gariazzo:2019gyi,Akita:2020szl} and references therein).  

While the approach has been successfully used in the standard cosmological scenario (see, e.g.,~\cite{Grohs:2015tfy,Gariazzo:2019gyi,Akita:2020szl,Drewes:2024wbw}), it has limitations when applying it to new physics scenarios. To understand this, let us assume that we inject neutrinos with high energy $E_{\nu,\text{max}}$ in the temperature range from $T_{\text{EM}} = T_{\text{ini}}$ to $T_{\text{EM}} = T_{\text{fin}}$. Generically, the computational time of the discretization approach, required to evolve the system during this temperature range, scales as (see Appendix~\ref{app:discretization-scaling})
\begin{equation}
    t_{\text{comp}} \propto E_{\nu,\text{max}}^{k+2}
    \label{eq:discretization-scaling}
\end{equation}
Here, the factor $k$ is the dimensionality of the collision integral after the analytic reduction of the integration:
\begin{equation}
\mathcal{I}_{\text{coll},\nu_{\alpha}} = \int \prod_{i = 1}^{k} d\xi_{i} F(\{\xi\}), 
\label{eq:analytic-reduction}
\end{equation}
with $\{\xi\} = \{\xi_{1},\dots,\xi_{k}\}$ being integration variables, and $F$ is some function depending on the distribution. The $k$ value is bounded from below by the standard cosmological scenario case, which is $k = 2$~\cite{Hannestad:1995rs}. New physics may drive the computational time~\eqref{eq:discretization-scaling} to enormously large values, or simply destroy the whole approach via spoiling the reduction~\eqref{eq:analytic-reduction}.

Indeed, first, the discretization approach requires simple analytic matrix elements in the neutrino source terms. In practice, this is not the case when we have hadronically decaying LLPs with mass $m\gg \Lambda_{\text{QCD}}$. This is because quarks and hadrons appearing in the decays undergo subsequent showering and hadronization. The latter results in a complicated phase space structure which is hard to fit in the form of an analytic matrix element.

Second, even if simple analytic matrix elements do exist, the computational complexity quickly increases if we depart significantly from the standard cosmological case. For example, simply increasing the integration dimensionality from $k =2$ to higher values may enormously increase the time of calculations. This is the case of, e.g., $2\to 3$ scatterings with neutrinos such as the famous $e^{+}e^{-}\to \nu\bar{\nu}\gamma$. Another example is when there are $n$-body decays with $n>3$, which are quite often for LLPs~\cite{Beacham:2019nyx}. 

Finally, the computational time problem exists even in the most optimistic case $k = 2$. Let us assume injections of neutrinos with large energy $E_{\nu}\gg T_{\text{EM}}$. They may appear in decays of heavy LLPs. Considering, e.g., $E_{\nu} \sim 1\text{ GeV}$ would enlarge the computational time compared to the standard cosmological case (where we assume $E_{\nu,\text{max}} = 20\text{ MeV}$) by a factor $\sim 50^{4} \sim 10^{7}$ (remind Eq.~\eqref{eq:complexity}), making any applications impossible in practice. Finally, depending on the energy density of the LLP, it may sizeably contribute to the Universe's energy density. For the same temperature range, the scale factor would be larger than in the Standard model case, which does not allow fixing the maximal comoving momentum in the grid $y_{\text{max}}$.

To summarize, there is no adequate approach to studying the dynamics of primordial neutrinos in the presence of new physics while maintaining model independence, efficiency, and transparency.

\section{Basics of DSMC}
\label{sec:DSMC-basics}
Consider the Liouville equation for the $N$-particle probability distribution density $F_{N}(\mathcal{R},\mathcal{V},t)$, where $\mathcal{R},\mathcal{V}$ is the set of coordinates and velocities of the particles, with a short-range potential $\Phi_{i,j}$ of binary interactions:
\begin{equation}
    \frac{\partial F_{N}}{\partial t} +\sum_{i = 1}^{N}\mathbf{v}_{i}\frac{\partial F_{N}}{\partial \mathbf{r}_{i}}+\sum_{1\leq i < j\leq N}\Phi_{i,j}F_{N} = 0
\label{eq:liouville}
\end{equation}
The DSMC approach approximately solves it using the following scheme (see~\cite{bird1970direct,ivanov1991theoretical,stefanov2019basic} and references therein):
\begin{itemize}
\item[1.] Apply the $N-1$ space variable reduction $F_{N}\to \tilde{F}_{N} = \int F_{N}\prod_{s = 2}^{N}d\mathbf{r}_{s}$. 
\item[2.] Switch to the iteration scheme by considering the equation on the time intervals $(t;t+\Delta t)$.
\item[2.] Decompose the space domain $\mathcal{D}$ onto disconnected sub-domains $\mathcal{D} =\cup_{l = 1}^{M} \mathcal{D}^{(l)}$ (``cells''), populated by fixed amounts of particles during $\Delta t$.
\item[3.] Split the evolution into three successive procedures within each time step: ballistic motion (free-streaming in the absence of collisions), binary collisions within each $\mathcal{D}^{(l)}$, and then interchanging particles between cells as a result of the first two steps. These collisions may change the kinematics of particles, their types, and number (e.g., via the collision $2\to n$).
\end{itemize}
Under an assumption that the system obeys ergodic conditions, the DSMC approach may be converted to an analog of the Bogoliubov–Born–Green–Kirkwood–Yvon hierarchy for $3+3N$ phase space, which reduces to the Boltzmann equation in the limit $N\to \infty$ and assuming the molecular chaos (i.e., that the velocities of colliding particles are statistically independent). 

\subsection{No-Time-Counter scheme}

The central part of the DSMC approach is to simulate the evolution of particles within an individual cell. There are various methods~\cite{bird1967velocity,koura1970transient,koura1986null,bird1989perception,roohi2016collision,stefanov2019basic}. The most efficient ones have $\mathcal{O}(N)$ computational complexity. Examples of the latter are No-Time-Counter (NTC), Majorant collision frequency, Simplified Bernoulli Trial, and others~\cite{roohi2016collision}. Here, we will discuss the NTC method, proposed in~\cite{bird1989perception}, which we will adapt for our purposes.

First, one defines the timestep of the simulation $\Delta t$. It must be sufficiently small to resolve the characteristic interaction time in the system. It may be calculated as
\begin{equation}
\Delta t = \left(\frac{(\chi_{\text{particle}}\cdot \sigma v)_{\text{max}} \cdot \mathcal{N}}{V_{\text{system}}}\right)^{-1}
\label{eq:timestep-DSMC}
\end{equation}
Here, $\mathcal{N}$ is the number of \textit{computational} particles (those actually used in the simulation), on the opposite of the number of \textit{physical} particles $N$. $\chi_{\text{particle}}$ is the particles' weight (see below); $V_{\text{system}}$ is the system's volume; $\sigma$ is the interaction cross-section; $v$ is the relative velocity. The subscript ``max'' denotes finding the maximal value among the system.

Let us discuss the relation between $N$, $\mathcal{N}$, and macroscopic observables. The quantity $N$ is fixed by the volume $V_{\text{system}}$ to represent the number density of the $i$th species, $n_{i} = N_{i}/V_{\text{system}}$. In its turn, it is related to the number of computational particles actually used in the simulation, $\mathcal{N}$, as $N = \sum_{i =1}^{\mathcal{N}}\chi_{i}$, where $\chi_{i}$ are individual weights of the particles. They need to be introduced if we address some redundancy in the system by replacing multiple particles with a single one. For example, in the setup without charge asymmetries, there is no need to consider particles and antiparticles separately. We can replace electrons and positrons with a single particle having the weight $\chi_{e} = 2$.

Next, consider splitting the system's volume into cells. Let us assume that there are $n_{\text{cells}}$ cells, each having the volume $V_{\text{cell}} = V_{\text{system}}/n_{\text{cells}}$. In the standard DSMC application cases, particular cells contain $N_{\text{cell}}\equiv \mathcal{N}/n_{\text{cells}}$ particles as low as $\mathcal{O}(10-20)$ and even lower, which is enough for simulating the evolution properly~\cite{stefanov2019basic,roohi2016collision}. Within a particular cell, one samples randomly 
\begin{equation}
N_{\text{sampled}} = \frac{N_{\text{cell}}(N_{\text{cell}}-1)}{2} \frac{\omega_{\text{cell,max}}\Delta t}{V_{\text{cell}}}
\label{eq:Nsampled}
\end{equation}
pairs of particles to interact. Here, $\omega_{\text{cell,max}} = (\chi_{\text{particle}}\sigma v)_{\text{cell,max}}$ is the estimate of the maximum interaction cross-section within the cell.

For each sampled pair, one accepts its interaction with the probability
\begin{equation}
P_{\text{acc}} = \frac{\omega}{\omega_{\text{cell,max}}}, \quad \omega = (\chi_{\text{particle}}\sigma v)_{\text{pair}}
\label{eq:interaction-acceptance}
\end{equation}
If the interaction is accepted, one simulates the possible final states for the given pair and its scattering kinematics.

The complexity of the NTC scheme grows as $\mathcal{O}(N_{\text{cell}})$~\cite{gallis2014direct}. This is achieved by the fact that $\omega_{\text{cell,max}}\Delta t/V_{\text{cell}}$ in the number of sampled events is typically $\ll 1$. The systems with the total number of particles $N\gg 10^{6}$ may be simulated within minutes, even on ordinary laptops. Such large values are already enough to reach the precision required in our studies.

The NTC method has been tested for various systems, including relativistic ones~\cite{schmidt2000new,wu2001analysis,wu2003parallel,venkattraman2012comparative,peano2009statistical}, which demonstrates its flexibility and coverage of the wide range of scenarios. 

\section{DSMC for neutrinos}
\label{sec:DSMC-neutrinos}

Let us now discuss how to apply the DSMC approach to study the evolution of primordial neutrinos.

As in the case of the state-of-the-art methods, we will first utilize the simplification coming from the properties of the Early Universe at the times of interest -- its homogeneity and isotropy. Because of this, we may drop the spatial degrees of freedom and treat the system as effectively zero-dimensional, with all interactions occurring at one point. Splitting the system into cells is a formal step to maintain performance because it allows parallelization for applying the NTC scheme. We will also neglect any cells' boundary interactions.

To accurately trace the thermalization of neutrinos, we represent their population by a set of individual particles characterized by the 4-momentum, flavor, and particle-antiparticle type. Every interaction involving neutrino (remind Eq.~\eqref{eq:neutrino-interactions}) would modify its properties. Namely, it may change its 4-momentum (if the interaction is elastic) and/or flavor (if it is the annihilation of the type $\nu_{\alpha}\bar{\nu}_{\alpha}\to \nu_{\beta}\bar{\nu}_{\beta}$). Finally, there are annihilation processes $\nu_{\alpha}\bar{\nu}_{\alpha}\leftrightarrow e^{+}e^{-}$, which may lead to a change in the number of neutrinos. 

\begin{figure}[h!]
    \centering
    \includegraphics[width=\linewidth]{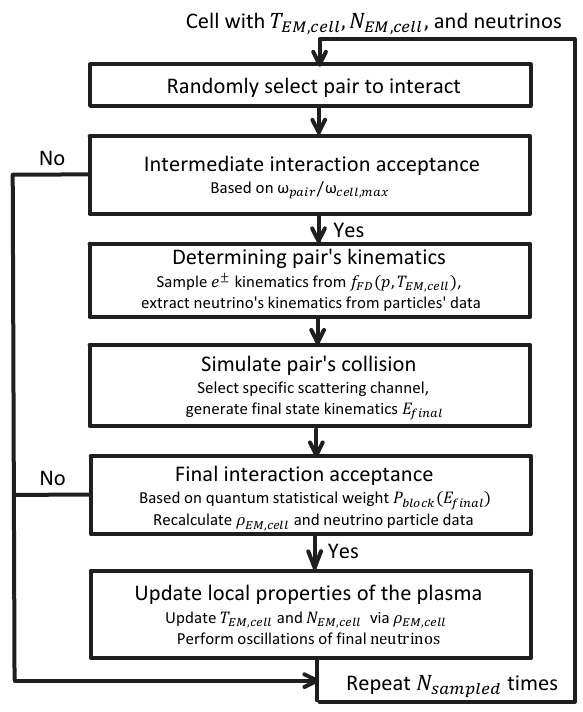}
    \caption{The modification of the No-Time-Counter scheme, used to simulate the interactions within the system's cells within the Direct Simulation Monte Carlo approach, for describing interactions in the MeV primordial plasma. First, we sample $N_{\text{sampled}}$ pairs to interact, Eq.~\eqref{eq:Nsampled}. For each pair, we compute its interaction weight and make an intermediate decision on whether it will interact using the criterion~\eqref{eq:interaction-acceptance}. Then, we sample the kinematics of the interacting particles, generate the final states resulting from the collision, and make the final decision of whether the interaction takes place from the Pauli principle~\eqref{eq:acceptance-pauli}. Finally, we update the local properties of the plasma: the EM plasma temperature and the number of EM particles, as well as neutrino flavor distributions by the oscillation probabilities, Eq.~\eqref{eq:oscillation-probabilities}.}
    \label{fig:new-NTC}
\end{figure}

Proceeding with the traditional DSMC method in the case of the primordial plasma with neutrinos is impossible, as it does not incorporate its fundamental features. These include the expansion of the Universe, the hierarchy between the equilibration rates in the neutrino and EM sectors, the Pauli principle, neutrino oscillations, and the presence of decaying particles. Below, we discuss these features and how we address them in detail (see also Fig.~\ref{fig:new-NTC}, showing the modification of the NTC scheme).

\begin{itemize} 
\item[1.] \textit{Expansion of the Universe}. From the DSMC's point of view, it simply represents an external force acting on the particles of the system, with an additional modification of increasing the system's (and cells') volume. These two effects may be simply accounted for by redshifting the total volume of the system $V_{\text{system}}$ (and hence the cell's volume) as well as the individual energies $E_{i}$ of the particles $\{i\}$, applied at each step of the simulation. Namely, at the beginning of the timestep $\Delta t$, we calculate the Hubble factor $H$ using Eq.~\eqref{eq:hubble}, and then make use of the relation
\begin{equation}
    V_{\text{system}} \to V_{\text{system}}(1+3H\Delta t), \quad E_{i} \to \frac{E_{i}}{1+H\Delta t}
\end{equation}
provided that $H\Delta t\ll 1$. To account for this requirement, we modify the definition~\eqref{eq:timestep-DSMC}:
\begin{equation}
   \Delta t = \text{min}\left[0.01 H^{-1}, \left(\frac{(\chi_{\text{particle}}\cdot \sigma v)_{\text{max}} \cdot N}{V_{\text{system}}}\right)^{-1} \right]
   \label{eq:timestep}
\end{equation}
Here, 0.01 is an arbitrary small factor. Given varying $H$, volume $V_{\text{system}}$, and neutrino energy $E_{\nu}$ throughout the evolution of the system, the value $\Delta t$ gets updated at the beginning of each iteration during the simulation using the following formula for the maximal rate:
\begin{equation}
    (\chi_{\text{particle}} \sigma v)_{\text{max}} = \frac{2 G_{F}^{2}}{3\pi}\cdot 2 E_{\nu,\text{max}} \langle E_{\nu}\rangle,
\end{equation}
where we used $\chi_{\text{particle}} = 1$ (provided that we implemented particles and anti-particles separately) and the estimate for $(\sigma)_{\text{max}}$ corresponding to the process $\nu_{\alpha}+\bar{\nu}_{\alpha}\to \nu_{\alpha}+\bar{\nu}_{\alpha}$, which is the fastest among all the neutrino interaction processes~\eqref{eq:neutrino-interactions}~\cite{Sabti:2020yrt}. The factor $2 E_{\nu,\text{max}} \langle E_{\nu}\rangle$ stays for the averaged invariant mass of the interacting high-energy neutrino with the rest of the neutrinos. 
\item[2.] \textit{Properties of the EM plasma}. As we discussed previously in Sec.~\ref{sec:neutrinos}, the reactions involving solely EM particles are orders of magnitude faster than those where neutrinos participate. As we have chosen the timestep $\Delta t$ comparable to the neutrino interaction rates, the EM particles may be viewed as a part of perfectly thermal plasma characterized by one parameter -- temperature $T_{\text{EM}}$. However, we then need to implement the response of any single interaction involving the EM particles on $T_{\text{EM}}$.

To reach this, at the beginning of the iteration, we characterize the EM plasma with the energy density $\rho_{\text{EM}}$, both globally (for the whole system) and locally (at the level of the individual cell). The global and cells' EM temperatures, which we denote by $T_{\text{EM}}$ and $T_{\text{EM,cell}}$ respectively, is related to the global and cell's energy densities of the EM plasma $\rho_{\text{EM}}$, $\rho_{\text{EM,cell}}$ by Eq.~\eqref{eq:rho-EM}.

During the NTC routine, the local number of electrons and positrons $N_{e^{\pm},\text{cell}}$ per cell is calculated from the relation between $T_{\text{EM,cell}}$ and the number density $n_{\text{EM}}(T_{\text{EM,cell}})$. The kinematics of any $e^{\pm}$ selected within the NTC algorithm is sampled from the Fermi-Dirac distribution $f_{\text{FD}}(p,T_{\text{EM,cell}})$. The change in $\rho_{\text{EM,cell}}$ resulting from the accepted interaction leads to the update in $T_{\text{EM,cell}}$ and $N_{e^{\pm},\text{cell}}$.

At the global level, once the simulations for all cells are performed, $\rho_{\text{EM,cell}}$s are merged into the total energy density $\rho_{\text{EM,system}}$, which allows obtaining the global temperature of the EM plasma.

\item[3.] \textit{Quantum statistics}. It enters the binary part of the collision integral~\eqref{eq:boltzmann-equation} with fermionic final states $F_{1},F_{2}$ having energies $E_{F_{1,2}}$ via multiplicative Pauli blocking factors 
\begin{equation}
P_{\text{block}} = (1-f_{F_{1}}(E_{F_{1}}))\times (1-f_{F_{2}}(E_{F_{2}})),
\label{eq:acceptance-pauli}
\end{equation}
where $f$ is the energy distribution of the given final state. Thus, it suppresses interactions where the final states would occupy the high-populated part of the energy distribution (e.g., $E\lesssim T$ for the equilibrium shape distribution with the temperature $T$). To implement this, one should consider the local energy distributions for both EM particles and neutrinos and calculate $P_{\text{block}}$. A possible simplification is, when calculating the blocking factor, to describe neutrino's distribution by the Fermi-Dirac function $f_{\text{FD}}(T_{\nu_{\alpha},\text{cell}})$, where $T_{\nu_{\alpha},\text{cell}}$ is the local effective neutrino temperature obtained in a way similar as we do for the EM plasma.\footnote{The actual neutrino distribution is, of course, non-thermal, and we use this approximation only when calculating $P_{\text{block}}$. Since the deviations from the thermality we study are not very large without loss of generality, we believe that the approximation is accurate.} 

\item[4.] \textit{Neutrino oscillations}. We incorporate them at the end of the iteration timestep by changing each of the neutrino flavors according to the formula
\begin{equation}
\nu_{\alpha}(E_{\nu}) 
\to \sum_{\beta}\langle P_{\alpha\beta}\rangle(E_{\nu},T_{\text{EM}})\nu_{\beta}(E_{\nu}), 
\label{eq:oscillation-probabilities}
\end{equation}
where $\langle P_{\alpha \beta}\rangle(E_{\nu},T_{\text{EM}})$ are averaged neutrino oscillation probabilities (remind also Eq.~\eqref{eq:boltzmann-equation}). For the neutrino oscillation parameters, we use the results from~\cite{Gonzalez-Garcia:2021dve}.

\item[5.] \textit{Presence of LLPs $X$ and new interactions}. Let us start with discussing LLPs. Further, we assume that LLPs are non-relativistic and decoupled at the temperatures of interest, which ideally matches the scope of this study. Decaying either into the EM plasma particles or neutrinos, they would heat the EM plasma temperature and distort the neutrino bath. 

Having the initial condition for the $X$'s number density, $n_{X}(T_{\text{ini}})$ for some temperature $T_{\text{ini}}$, at the beginning of the simulation, we add the amount $N_{X}$ of $X$ particles fixed in a way such that $N_{X}/V_{\text{system}} = n_{X}(T_{\text{ini}})$. Per each timestep $\Delta t$, provided that it is much smaller than the LLP's lifetime $\tau_{X}$, their number is evolved by the exponential distribution: $dN_{X}/dt = -N_{X}/\tau_{X}$.

For each decay, it is possible to obtain the energies of resulting neutrinos and calculate the amount of the EM energy using Monte-Carlo simulations -- the baseline approach for particle physics. This is a natural choice if one wants to maintain the model independence, as it is maximally general and may describe any process. In particular, exclusive decays (where we have well-defined ``fixed'' final states, e.g., $X\to 3\pi$) may be simulated on-flight by sampling the phase space of decay products using the analytic matrix element of the process. The phase space of hadronic decays in the LLP mass range $m\gg \Lambda_{\text{QCD}}$ (such as $X\to q\bar{q}\nu$, where $q$ is a quark) may be obtained by simulating them in \texttt{PYTHIA8}~\cite{Bierlich:2022pfr} for a grid of masses and subsequently using the output particle's data in the form of events inside the DSMC code.

The Monte Carlo sampler must incorporate the interactions of the decay products with the primordial plasma, which may substantially redistribute their energy between the neutrino and EM sectors compared to the vacuum case. Namely, all electrically charged particles with lifetimes $\tau \gtrsim 10^{-10}\text{ s}$, such as muons, charged pions, and kaons, appearing in the MeV plasma may undergo kinetic energy loss via EM interactions, annihilation, interactions with nucleons before decaying~\cite{Akita:2024ta}. This evolution may again be implemented probabilistically, in the spirit of Monte Carlo simulations. 

Absolutely similarly, it is possible to sample the energies for non-standard scattering processes, e.g., for the $2\to 3$ scatterings $e^{+}e^{-}\to \nu_{\alpha}\bar{\nu}_{\alpha}\gamma$.
\end{itemize}

In order to finish the discussion of the approach, let us address the question of the number of particles per cell, $N_{\text{cell}}$, entering Eq.~\eqref{eq:Nsampled}. In our system, it is
\begin{equation}
     N_{\text{cell}} = N_{e^{\pm},\text{cell}} + 2\sum_{\alpha}N_{\nu_{\alpha},\text{cell}}
\end{equation}
Since statistical quantities, such as temperatures, are involved in simulating the interactions, it is not possible to use small $N_{\text{cell}}\sim 10$, as it is typically done in the DSMC simulations. Instead, the values as large as $N_{\text{cell}} = \mathcal{O}(100)$ should be considered (see Appendix~\ref{app:convergence-tests}). As a bonus, such a large number also allows for avoiding various stochastic problems of the NTC method, including repeated interaction of the same pair~\cite{ROOHI20161}.

Now, let us discuss the values of $N,N_{\text{cell}}$ we use and the scaling of the DSMC simulation time with the maximal neutrino energy in the system $E_{\nu,\text{max}}$. More details may be found in Appendix~\ref{app:convergence-tests}, and here we make a summary.

The typical number of particles per neutrino flavor we consider in the setup is $N_{\nu} 
\simeq 10^{6}$, which results in $N= \text{few}\times 10^{6}$. The standard number of particles per cell we have chosen is $N_{\text{cell}}=400$. These numbers are enough to keep the statistical noise at the level of $0.1\%$ in the absence of high-energy neutrinos.

The scaling of the computational time with $E_{\nu,\text{max}}$ is \textit{linear} to \textit{quadratic} (in some marginal cases, as we comment on below). The scaling comes from the unavoidable linear dependence of the number of timesteps on $E_{\nu,\text{max}}$ (remind Eq.~\eqref{eq:timestep}) and the possible scaling $N(E_{\nu,\text{max}})$. The latter may be required to maintain a computationally large enough number of high-energy injected neutrinos to avoid fluctuations in the microscopy of thermalization. As far as $E_{\nu,\text{max}}\lesssim 1\text{ GeV}$, $N$ may be kept constant, and the scaling of the simulation time is linear. If $E_{\nu,\text{max}}\gtrsim 10\text{ GeV}$, one would need to increase $N$ to keep the number of injected neutrinos large enough to avoid fluctuations. However, the increase is \textit{linear} with $E_{\nu,\text{max}}$. As a result, in this worst-case scenario, the scaling of the running time becomes $E_{\nu,\text{max}}^{2}$ -- still much better than the scaling of the discretization approach, Eq.~\eqref{eq:discretization-scaling}.\footnote{There may be, in principle, an additional slowdown coming from the need to distribute particles into cells at the beginning of each iteration. We have checked that this splitting only costs a tiny fraction of the whole time independently of the value of the number of computational particles $\mathcal{N}$ and, hence, does not add anything on top of the expected scaling.}

\section{Current implementation}
\label{sec:current-implementation}
We have implemented a simplified version of the DSMC method described above, which serves as proof-of-principle.\footnote{The code may be provided upon request.} 

The main approximation of the \textit{current} implementation is that we have neglected the electron mass $m_{e}$ when describing the population of the EM particles; this is done to simplify the sampling of $e^{\pm}$ particles and relate the total energy of the EM plasma to its temperature. 

Although keeping $m_{e}$ finite is necessary to know the final value of neutrino-to-EM energy densities ratio ($=N_{\text{eff}}$), it is irrelevant for studying the main topic of this work -- non-equilibrium dynamics of neutrinos at the times when they start decoupling, and qualitative behavior such as the sign of the correction $\Delta N_{\text{eff}} = N_{\text{eff}} - N_{\text{eff}}^{\Lambda\text{CDM}}$. This is because of two reasons. First, $m_{e}$ does not affect the dynamics at MeV temperatures (the domain of interest of this study), since it can be simply neglected compared to the typical electrons' energies of $E_{e} \approx 3.15 \cdot T_{\text{EM}}$. To validate this statement, we compare the predictions of DSMC at MeV temperatures with the approaches keeping finite electron mass and NLO QED corrections, and find a perfect agreement. Second, including it at lower temperatures cannot change the sign of $\Delta N_{\text{eff}}$, modifying only its magnitude $\Delta N_{\text{eff}}$. 

We will include the electron mass and its QED corrections~\cite{Heckler:1994tv,Fornengo:1997wa,Bennett:2019ewm} in future papers delivering the full DSMC implementation.

The implementation is written in \texttt{Mathematica}. However, low-level routines, such as simulations of interactions and manipulations with cells, are compiled in \texttt{C++}. This approach allows combining moderate performance with symbolic calculations, which are needed when dealing with describing kinematics and deriving the matrix elements of various processes. Also, it makes it possible to use existing realizations of Monte-Carlo sampling of decays of LLPs, such as \texttt{SensCalc}~\cite{Ovchynnikov:2023cry}. In the next revisions, we will write a part of the code in native \texttt{C++} and use it as a library inside \texttt{Mathematica}.

On a laptop with CPU \texttt{AMD Ryzen AI 9 HX 370}, the running time required to produce most of the plots below is $\lesssim 5$ minutes; it varied only mildly depending on the setup, including the energies of the neutrinos included in the system. In particular, in order to produce the neutrino distributions shown in Fig.~\ref{fig:injection-70-MeV-5-perc}, we spent only $30$ seconds. We expect significant improvement, possibly by an order of magnitude, in the running time after optimizing the code and/or rewriting some of its modules in native C++. Finally, with the implementation, we maintain the approximate linear scaling of the computational time with $N$, as expected from the basics of the NTC approach.

To validate the developed neutrino DSMC, we have studied its predictions in the case of well-established scenarios, including:
\begin{enumerate}
    \item \textit{Approaching thermal equilibrium.} In the absence of Universe expansion, independently of the initial conditions, neutrinos have to reach thermal equilibrium with the EM particles. In particular, their differential distribution in the number and energy densities, which we will plot throughout the paper, must be
    \begin{align}
    \frac{dn_{\nu}}{dE_{\nu}} =& \frac{g_{\nu}}{2\pi^{2}}f_{\text{FD}}(E_{\nu},T_{\nu})\times E_{\nu}^{2}, \\ 
        \frac{d\rho_{\nu}}{dE_{\nu}} =& \frac{g_{\nu}}{2\pi^{2}}f_{\text{FD}}(E_{\nu},T_{\nu})\times E_{\nu}^{3},
        \label{eq:rho-distribution-thermal}
    \end{align}
    where $T_{\nu} = T_{\text{EM}}$ is the neutrino temperature, two powers of $E_{\nu}$ come from the phase space, and one in Eq.~\eqref{eq:rho-distribution-thermal} from the definition of $\rho_{\nu}$. Finally, $f_{\text{FD}}$ is the Fermi-Dirac distribution (remind Eq.~\eqref{eq:f-FD}), with $g_{\nu}$ being the lepton charge degree of freedom (remind Eq.~\eqref{eq:Hubble-g}).
    
    Eq.~\eqref{eq:rho-distribution-thermal} automatically implies that in equilibrium, the ratio of the energy densities of the neutrino and EM plasmas is
    \begin{equation}
        \left(\frac{\rho_{\nu}}{\rho_{\text{EM}}}\right)_{\text{eq}} = \frac{7/8 \cdot g_{\nu}}{7/8\cdot g_{e} + g_{\gamma}} = \frac{21}{22},
         \label{eq:rho-ratio-thermal}
    \end{equation}
    where we have used Eq.~\eqref{eq:rho-EM} and assumed $T_{\text{EM}}\gg m_{e}$.
    \item  \textit{Energy transition rates.} Consider the initial setup where the distribution function of neutrinos is fixed by $f_{\text{FD}}$, parametrized with the temperature $T_{\nu_{\alpha}}\neq T_{\text{EM}}$. During the equilibration and in the absence of expansion, the energy transition rates between the neutrino and EM sectors must match the well-known analytic result from~\cite{EscuderoAbenza:2020cmq} (where we turn off the expansion as well). 
\item \textit{Expansion and decoupling.} If including the expansion of the Universe in the previous setup, we should consistently recover the decoupling of neutrinos, which prevents their population from full thermalization, as well as reproduce the results of~\cite{EscuderoAbenza:2020cmq}.
\end{enumerate}

Details may be found in Appendix~\ref{app:cross-checks}. In addition, we have performed tests that are not present in the paper. Those include the evolution of neutrinos and antineutrinos (the evolution must preserve the lepton symmetry up to Monte Carlo fluctuations) and independence on the exact simulation setup (\textit{e.g.} Number of simulation cells, the total number of particles, etc.). We believe that it proves that our approach fulfills the requirements to be accepted as a valid method for treating the evolution of neutrinos.

\section{Case studies} 
\label{sec:applications}
To demonstrate the potential of various implications of the DSMC method, we will consider several toy case studies specified by the initial conditions on the neutrino distribution functions. These setups have two applications. On the one hand, they mimic distinct scenarios with new physics and thus provide useful insights into the dynamics of the primordial plasma. On the other hand, they will comprehensively demonstrate the performance and flexibility of the neutrino DSMC approach. 

Firstly, we investigate the evolution of a system where neutrinos initially possess an equilibrium energy distribution with a temperature $T_{\nu_{\alpha}} \neq T_{\text{EM}}$ (see Sec.~\ref{sec:application-T}). This setup encompasses two distinct scenarios. The first scenario arises when energy is injected exclusively into the electromagnetic (EM) sector, resulting in $T_{\text{EM}} > T_{\nu_{\alpha}}$. The second scenario occurs when nearly thermal neutrinos are introduced into the neutrino sector, as explored in~\cite{EscuderoAbenza:2020cmq}. These cases can be analyzed using the integrated Boltzmann equation developed in~\cite{Escudero:2018mvt,EscuderoAbenza:2020cmq}. Nevertheless, we will demonstrate that even within these simplified setups, deviations from the thermal shape of the neutrino distribution emerge, leading to discrepancies between the solutions of the unintegrated and integrated approaches to the neutrino Boltzmann equation, particularly in the determination of $N_{\text{eff}}$.

Second, we will consider injections of high-energy monochromatic neutrinos (Sec.~\ref{sec:application-instant}). This scenario represents the case of two-body decays of heavy LLPs, such as neutrinophilic scalars~\cite{Kelly:2019wow}, majorons~\cite{Kim:1986ax}, $B-L_{\alpha}$ mediators~\cite{Ilten:2018crw}, and relics in late reheating scenarios~\cite{Hannestad:2004px}. In the context of our studies, it is preferable over the realistic continuous injections by decaying LLPs because of the transparency of the analysis; despite the simplicity, understanding the dynamics of instant injections provides qualitative insights for the continuous decays, which we explore in our companion paper~\cite{Ovchynnikov:2024xyd}.\footnote{For the instant injections, the computational time of both the discretization approach and DSMC does not include scaling of the number of timesteps with the injected neutrino energy $E_{\nu,\text{max}}$ (remind Appendix~\ref{app:discretization-scaling}). This is because they quickly lose their energy, and the timestep required to resolve their thermalization no longer depends on $E_{\nu,\text{max}}$.}

We will consider high injection temperatures, $T_{\text{EM}}\gtrsim 1\text{ MeV}$. We will show that in the case of sufficiently large neutrino energy, such that $E_{\nu}\gg T_{\text{EM}}$, these injections would result in a decrease in the neutrino-to-EM energy densities ratio compared to the standard cosmological scenarios. This setup will also serve to demonstrate that the performance of the DSMC does not depend on the neutrino energy (supporting the initial expectations) and to cross-check it by comparing the neutrino evolution with the predictions of the discretization codes.

Finally, we will study injections of neutrinos from decays of different long-lived SM particles, such as muons, charged pions, and kaons (Sec.~\ref{sec:application-mesons}). This case corresponds to a common scenario of LLPs with complex decay chains, which may not decay into neutrinos directly but instead decay into such heavy states. Examples are, e.g., a decay of the Higgs-like scalars into $\pi^{+}\pi^{-}/K^{+}K^{-}$, the dark photon decay into $2\pi/3\pi/4\pi$, and decays of HNLs into $\pi \mu$~\cite{Bondarenko:2018ptm}. Another illustrative case is the decay into quarks, where we have a high multiplicity of meson states. We will show that independently of the decaying particle (or the fraction of their energy placed to the neutrino plasma right after decay), the ratio~\eqref{eq:delta-rho} decreases below the equilibrium value. This case also study demonstrates the flexibility of our approach, which may handle any decay chain with complicated kinematics.

To make the illustrative analysis for this and other studies performed in this paper, we introduce the quantity
 \begin{equation}                
\delta \rho_{\nu} =\left(\frac{\rho_{\nu}}{\rho_{\text{EM}}}\right)^{-1}_{\text{eq}}\frac{\rho_{\nu}}{\rho_{\text{EM}}} - 1
\label{eq:delta-rho}
\end{equation}

Throughout the section, we will compare the predictions of DSMC with the modified discretization approach from~\cite{Akita:2020szl}, which we develop for the work~\cite{Akita:2024ork}. The brief description of the approach can be found in Appendix~\ref{app:discretization-approach}. It serves two important purposes. First, the overall agreements between DSMC and this approach serve as a very robust cross-check of our method, showing how well it traces the dynamics of neutrinos. In particular, in the discretization approach, we keep finite electron mass and include LO QED corrections. Finding the very good agreement, we validate our approximation of neglecting the electron mass in our proof-of-principle study. 

Second, we will compare the performances of the two methods, highlighting the setups where the discretization approach becomes inapplicable in practice.

\subsection{From equilibrium spectral shapes to distortions}
\label{sec:application-T}
Let us consider a system with neutrinos having an equilibrium shape of energy distributions, but the temperatures of these distributions differ from the EM plasma temperature. 

We will study how the equilibration of this initial condition evolves in time, to identify the possible deviations from the description dynamics of the equilibration following Ref.~\cite{EscuderoAbenza:2020cmq}, where we turn off the electron mass in order to compare apples with apples. These deviations genuinely appear from the non-thermal distortions in the neutrino sector (invisible within the method of~\cite{EscuderoAbenza:2020cmq}). It is because the interaction rates of different parts of the neutrino spectrum are energy-dependent (remind Sec.~\ref{sec:neutrinos}). 

We will consider the particular initial condition where neutrinos have the same temperature $T_{\nu_{\alpha}} = 3.5\text{ MeV}$, and the EM plasma has a lower temperature $T_{\text{EM}} = 3\text{ MeV}$. 

\begin{figure}[t!]
    \centering
    \includegraphics[width=0.9\linewidth]{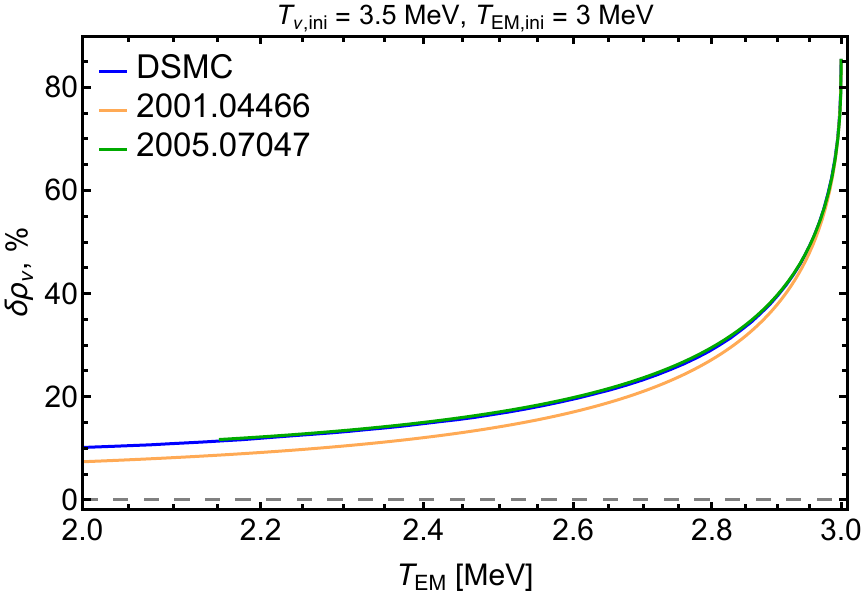} \\ \includegraphics[width=0.9\linewidth]{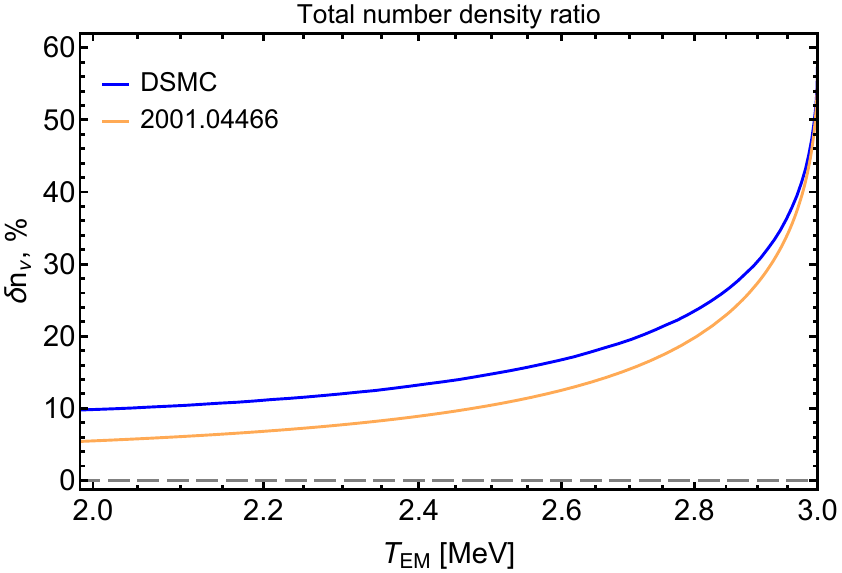}  
    \caption{The evolution of neutrinos and the EM plasma energy densities ratio under the scenario where the neutrino distribution shape is thermal (Eq.~\eqref{eq:distribution-thermal}), but has temperature $T_{\nu}$ different from the EM plasma $T_{\text{EM}}$. For the initial setup, we consider $T_{\nu}= 3.5\text{ MeV}$ and $T_{\text{EM}} = 3\text{ MeV}$. \textit{Top panel}: The energy densities ratio $\delta \rho_{\nu}$, given by Eq.~\eqref{eq:delta-rho}. The blue line shows the result of our DSMC approach, the green line denotes the prediction of the modified code from~\cite{Akita:2020szl,Akita:2024ork}, whereas the orange line is obtained using the method of integrated neutrino Boltzmann equations from~\cite{EscuderoAbenza:2020cmq}, which assumes that the shape of the neutrino distribution is perfectly thermal throughout the whole evolution. \textit{Bottom panel}: The analog of $\delta \rho_{\nu}$ but for the number densities of neutrinos and the EM particles, highlighting the deviation from the thermality of the neutrino spectrum throughout the evolution.}
    \label{fig:delta-rho-case-study-1}
\end{figure}

The resulting evolution of $\delta \rho_{\nu}$, as predicted by the DSMC approach and the method from~\cite{EscuderoAbenza:2020cmq}, is shown in Fig.~\ref{fig:delta-rho-case-study-1}. From the figure, we see that in terms of $\delta \rho_{\nu}$, the two descriptions match at the initial stages, while the deviations appear once the system develops, signaling the accumulating neutrino spectral distortions. They get frozen throughout the evolution because of the decoupling of neutrinos. The same conclusion holds in the opposite case of the initial condition $T_{\text{EM}}>T_{\nu}$. The accumulation of the distortions is easily visible if considering the analog of $\delta \rho_{\nu}$ but for the number density of neutrinos, which we show in the bottom panel of the figure. It demonstrates that for this setup, the distortions build up because of the suppression of the $\nu\bar{\nu}\to e^{+}e^{-}$ annihilation rate, which keeps the extra energy stored in the neutrino sector.

Therefore, we conclude that the integrated Boltzmann approach may provide insufficient accuracy even in cases where there are no direct distortions of the neutrino spectrum (see further discussion of this point in Ref.~\cite{Akita:2024ork}).

The DSMC predictions perfectly agree with the discretization method from~\cite{Akita:2020szl,Akita:2024ork}. Both approaches work reasonably fast -- within a minute, but the discretization approach is $\mathcal{O}(2)$ times faster. This is explained by the smallness of the maximal neutrino energy -- the scenario for which the discretization works well.

\subsection{Instant neutrino injection}
\label{sec:application-instant}

Let us proceed to a different scenario in which there are injections of non-thermal neutrinos with $E_{\nu}\gg T_{\text{EM}}$. For this setup, the integrated Boltzmann approach is completely inapplicable, as high-energy neutrinos have a much larger rate of interactions than their thermal counterparts, severely influencing the dynamics of the thermalization even if their amount is low.

We will study the injection of monochromatic neutrinos with energy $E_{\nu,\text{inj}}$ at temperature $T_{\text{EM}} = 3\text{ MeV}$, and consider three different values $E_{\nu,\text{inj}} = 20, 70, 500 \text{ MeV}$. We will analyze both the evolution of $\delta \rho_{\nu}$ and the neutrino spectrum shape. 

The option $E_{\nu,\text{inj}} = 20\text{ MeV}$ primarily serves to demonstrate the necessity of using the unintegrated Boltzmann approach in case of non-thermal distortions. The second setup $E_{\nu,\text{inj}} = 70\text{ MeV}$ is central -- it will show the qualitative impact of large neutrino energies on $N_{\text{eff}}$. We will use it to compare with the discretization codes from~\cite{Boyarsky:2021yoh,Akita:2020szl,Mastrototaro:2021wzl}, predicting contradictive behavior of the sign of $N_{\text{eff}}-N_{\text{eff}}^{\Lambda\text{CDM}}$ in presence of high-energy neutrinos. Finally, the highest energy case $E_{\nu,\text{inj}} = 500\text{ MeV}$ highlights the performance of our setup -- the running time and precision are almost independent of the neutrino energy.  

\subsubsection{Injection of 20 MeV neutrinos}

\begin{figure}
    \centering
    \includegraphics[width=0.45\textwidth]{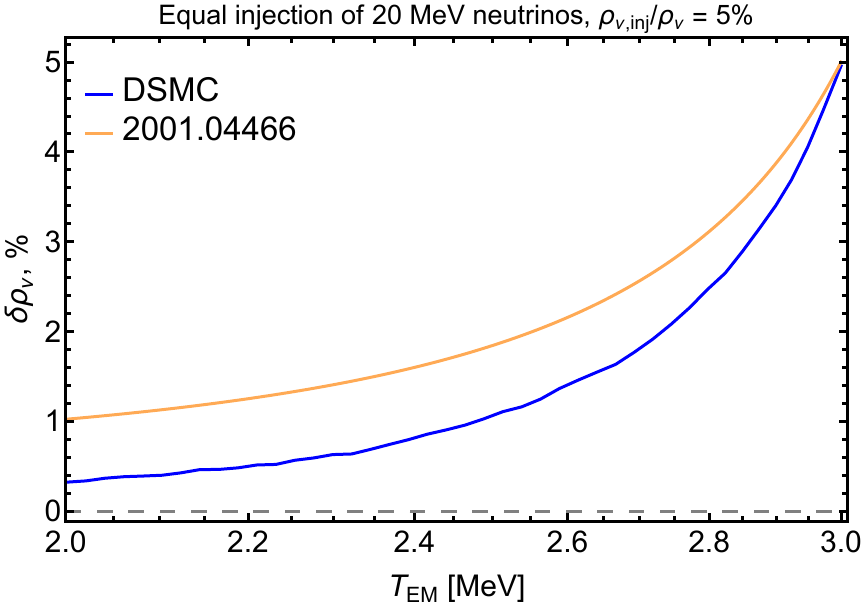}
    \caption{The behavior of the ratio~\eqref{eq:delta-rho} under the injection of 20 MeV neutrinos equally to all neutrino flavors at the temperature $T_{\text{EM}} = 3\text{ MeV}$. The total injected energy density is $\rho_{\nu,\text{inj}}/\rho_{\nu,\text{total}} = 5\%$. The blue line shows the prediction of the DSMC method, whereas the orange one corresponds to the integrated Boltzmann approach from~\cite{EscuderoAbenza:2020cmq}.}
    \label{fig:injection-20-MeV}
\end{figure}
Consider the injection of 20 MeV neutrinos. We assume equal injection among the three neutrino flavors, with the total injected energy density $\rho_{\nu,\text{inj}}/\rho_{\nu,\text{total}} = 5\%$. Here and below, we include the Hubble expansion of the Universe, but turn off the neutrino oscillations. 

The evolution of the resulting $\delta \rho_{\nu}$ is shown in Fig.~\ref{fig:injection-20-MeV}, where we, as usual, also include the prediction of the integrated Boltzmann approach. Both approaches predict a monotonic decrease of $\delta \rho_{\nu}$. In particular, at late temperatures, when the expansion prevents equilibrating, we end up with the value of $\delta \rho_{\nu}$ close to $0$. However, the rate of decrease of $\delta \rho_{\nu}$ predicted the neutrino DSMC is much faster. This is explained by the fact that, compared to thermal particles, the injected high-energy neutrinos have a larger probability of interacting with the EM sector and, hence, transporting their energy.

\subsubsection{Injection of 70 MeV neutrinos}

Let us now proceed with the 70 MeV injection. We will consider several setups here. The first one is with equal injection among the neutrino flavors and a large $\rho_{\nu,\text{inj}}/\rho_{\nu,\text{total}} = 30\%$. It serves as a very illustrative demonstration of the qualitative features of the evolution of $\delta \rho_{\nu}$. The two others are with the smaller injected energy $\rho_{\nu,\text{inj}}/\rho_{\nu,\text{total}} = 5\%$ and two different injection patterns: equal energy distribution among the flavors, and the injection solely into the sector of electron neutrinos. We will use them to compare with the predictions of different discretization codes from the literature.

\begin{figure}[h!]
    \centering
    \includegraphics[width=0.45\textwidth]{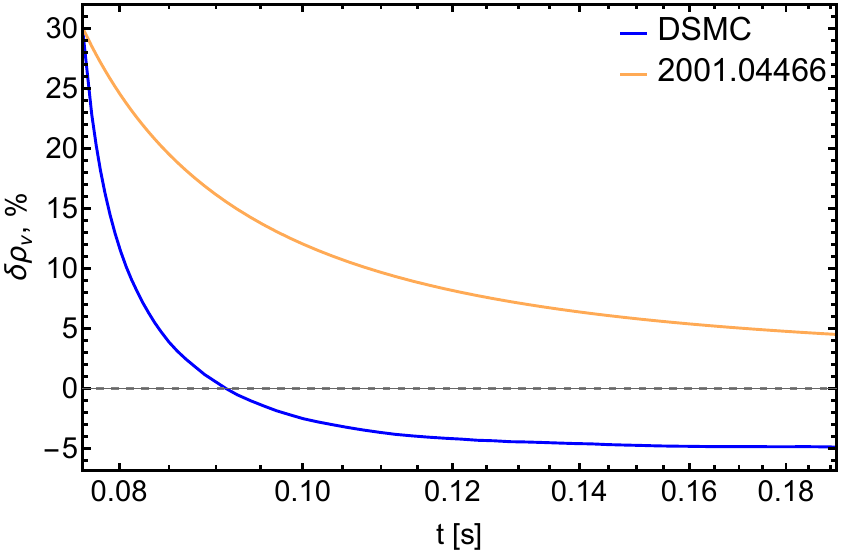}
    \\ \includegraphics[width=0.45\textwidth]{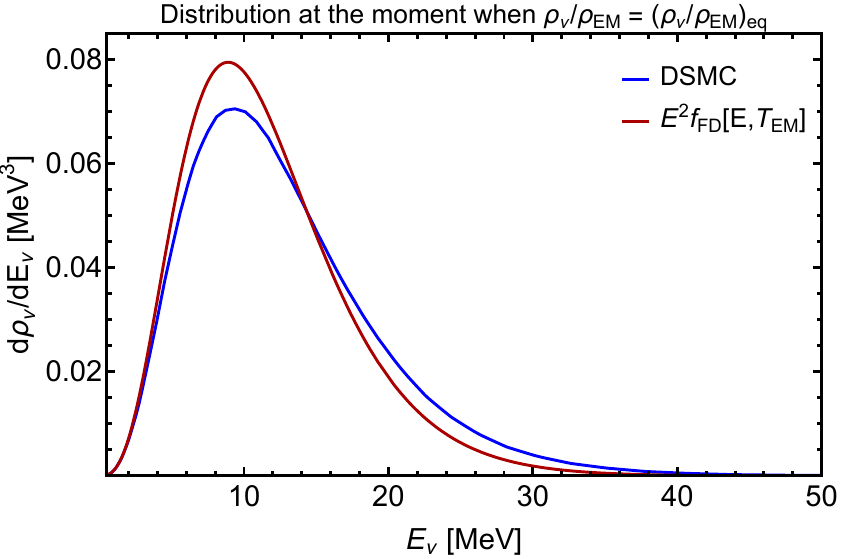}
    \caption{The temporal evolution of the plasma after the injection of neutrinos with energies $E_{\nu} = 70\text{ MeV}$ and the overall energy density $\rho_{\nu,\text{inj}}/\rho_{\nu,\text{total}} = 30\%$. The other parameters of the setup are similar to the one considered in Fig.~\ref{fig:injection-20-MeV}. \textit{Top panel}: the behavior of $\delta \rho_{\nu}$ with temperature, where we show the predictions of the DSMC (the blue curve) and the integrated approach from Ref.~\cite{EscuderoAbenza:2020cmq} (the orange curve). \textit{Bottom panel}: comparison of the shape of the neutrino energy distribution for the system from Fig.~\ref{fig:injection-70-MeV-30-perc} at the moment when $\delta \rho_{\nu} = 0$ during the equilibration, as obtained with the DSMC simulation (the blue curve) and assuming the equilibrium neutrino spectrum (the red curve).}
    \label{fig:injection-70-MeV-30-perc}
\end{figure}

\begin{figure*}[t!]
    \centering
    \includegraphics[width=0.4\textwidth]{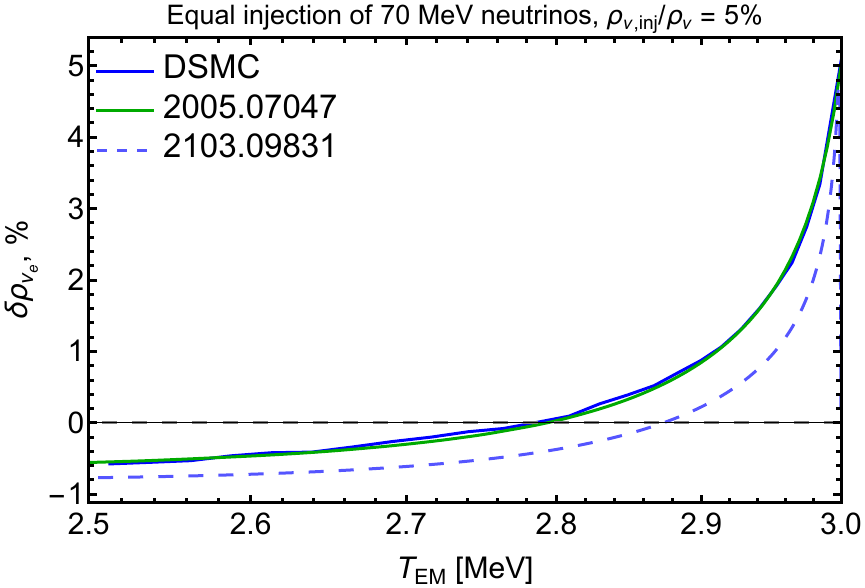}~\includegraphics[width=0.45\textwidth]{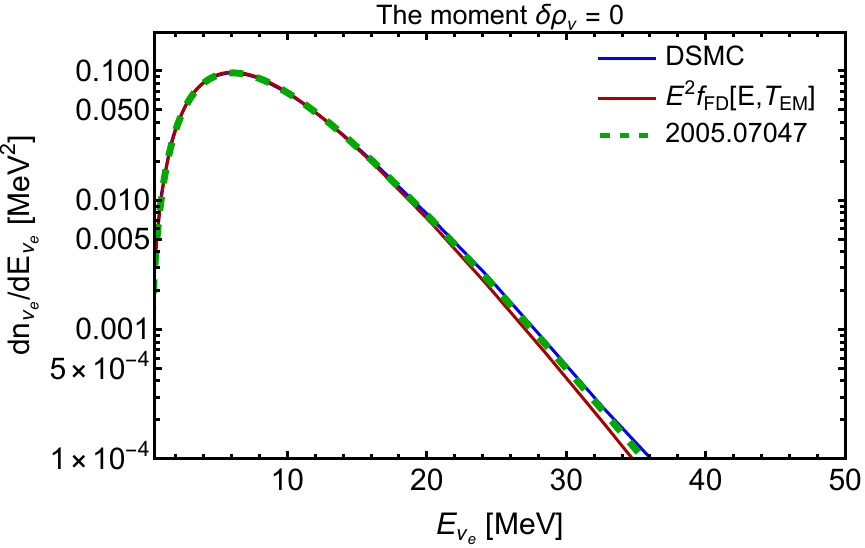} \\ \includegraphics[width=0.4\textwidth]{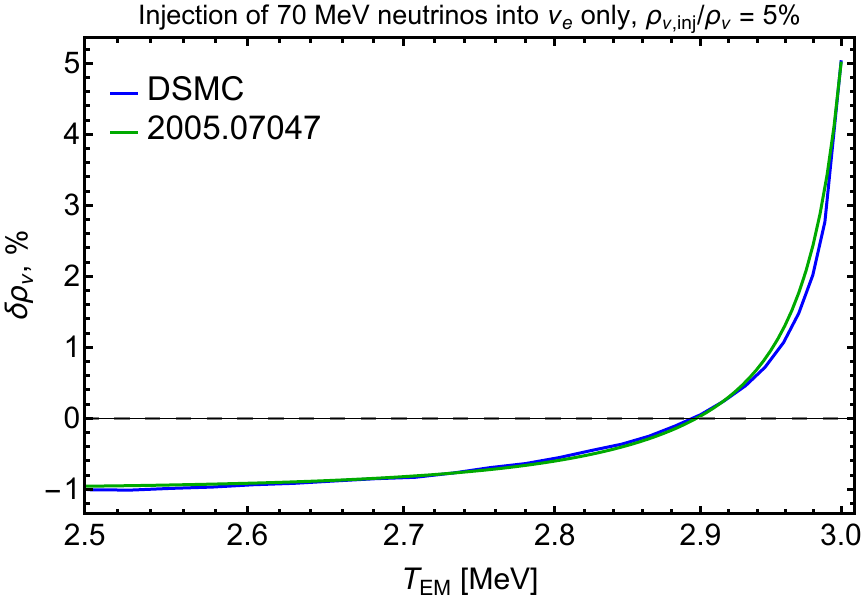}~\includegraphics[width=0.45\textwidth]{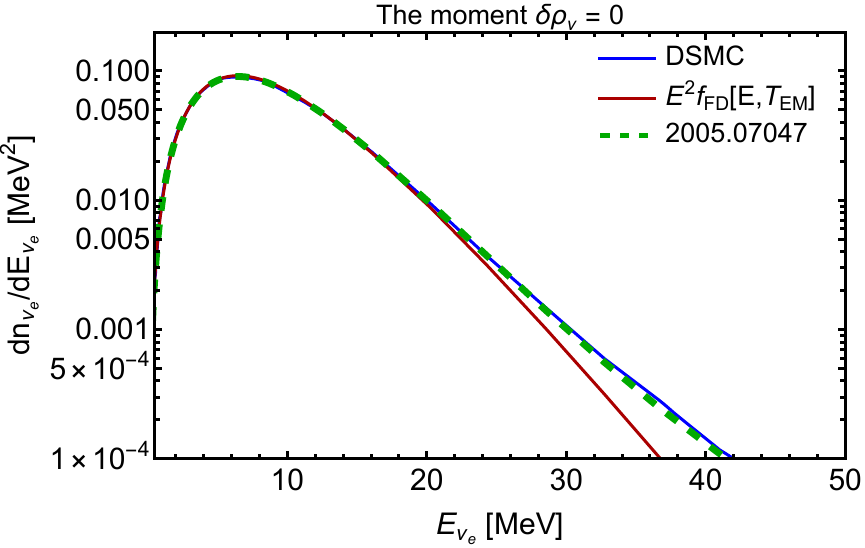}
    \caption{Comparison of the DSMC approach with the discretization codes for the setup of injection of 70 MeV neutrinos at $T_{\text{EM}} = 3\text{ MeV}$. Two configurations are considered: equal injection among the flavors (the top panels) and the injection solely into $\nu_{e}$ (the bottom panels). In both cases, the injected energy fraction is $\rho_{\nu,\text{inj}}/\rho_{\nu,\text{total}} = 5\%$. The left plots show the evolution of $\delta \rho_{\nu}$, given by Eq.~\eqref{eq:delta-rho}.  In the plots, the blue lines are the DSMC predictions, the green lines denote the calculation by the discretization approach from~\cite{Akita:2020szl} (see also~\cite{Akita:2024ta}), the dashed blue line is the result obtained in~\cite{Boyarsky:2021yoh} (see text for discussions). The right plots are snapshots of the electron neutrino distribution spectrum at the temperature when $\delta \rho_{\nu} = 0$. In addition to the results from the DSMC and Ref.~\cite{Akita:2024ork}, we include the plot of the equilibrium neutrino distribution given by $E_{\nu}^{2}f_{\text{FD}}(E_{\nu},T_{\text{EM}})$ (solid red lines).}
    \label{fig:injection-70-MeV-5-perc}
\end{figure*}
Fig.~\ref{fig:injection-70-MeV-30-perc}, upper panel, shows the evolution of $\delta \rho_{\nu}$ for the 30\% injection setup. Now, there is a qualitative difference in its behavior between the integrated and DSMC approaches. The former results in the naively expected monotonic decrease of $\delta \rho_{\nu}$, whereas according to the latter, it first rapidly drops below zero, where it then freezes out. Without the expansion of the Universe, it would have been a decrease of $\delta \rho_{\nu}$ to negative values, and then a slow monotonic reaching $\delta \rho_{\nu} \to 0$ from below.

To understand this counter-intuitive result, let us remind Sec.~\ref{sec:neutrinos} and highlight two important properties of the plasma: \textit{(i)} EM particles instantly equilibrate between themselves, and \textit{(ii)} weak interaction rates grow with the invariant mass of colliding particles. Because of this, the injected non-thermal neutrinos quickly ``knock out'' thermal neutrinos by the interactions 
\begin{equation}
\nu_{\text{inj}}
\bar{\nu}_{\text{thermal}}\to e^{+}e^{-}, \quad \nu_{\text{inj}}
\nu_{\text{thermal}}\to \nu\nu
\end{equation}
The first process pumps the injected energy and a fraction of the energy of the thermal population to the EM sector. The rate of these processes is much higher than the rate of the same processes when only thermal particles are involved. Knocking out thermal neutrinos determines the shape of the neutrino spectrum during these interactions: compared to the equilibrium spectrum $f_{\text{FD}}$, it is underabundant in small energies and overabundant in large energies. 

The snapshot of the neutrino spectrum at the moment when $\delta \rho_{\nu} = 0$ is shown in the lower panel of Fig.~\ref{fig:injection-70-MeV-30-perc}. Then, we have equilibrium amounts of energies in the EM and neutrino sectors. However, while the EM plasma has a perfect thermal spectrum, the neutrino spectrum has a shift to higher energies. 

The further dynamics of $\delta \rho_{\nu}$ depends on the balance between the energy transfer rates $\nu\to \text{EM}$ and $\text{EM}\to \nu$. Because of the energy dependence of the weak processes' rate, the overabundance of the high-energy neutrino leads to the faster transfer $\nu\to\text{EM}$ than $\text{EM}\to \nu$, where we have thermal electrons. As a result, $\delta \rho_{\nu}$ continues falling below zero until neutrino-induced heating of the EM plasma temperature and/or the expansion of the Universe turn the negative energy transfer from the neutrino sector to zero. 

Since the sign of $\delta \rho_{\nu}$ is associated with the sign of the correction $\Delta N_{\text{eff}}$, we conclude that the injection of such high-energy neutrinos is associated with a decrease in $N_{\text{eff}}$ below its $\Lambda$CDM value. This conclusion holds in the case when the EM plasma temperature is high enough during the neutrino injection, such that the interactions between the neutrinos and the EM plasma are possible.

A similar result has been obtained in our previous work~\cite{Boyarsky:2021yoh}, which considered a setup with the injection of 70 MeV neutrinos but with a smaller amount within the discretization approach. The same behavior has been observed when considering the cosmological impact of HNLs decaying mainly into neutrinos (see also Refs.~\cite{Ruchayskiy:2012si,Rasmussen:2021kbf}). These results, however, contradicted Ref.~\cite{Mastrototaro:2021wzl} (see also~\cite{Dolgov:2000jw}), which studied the same setup with HNLs with masses below the pion mass and found that $N_{\text{eff}}$ may only increase. Given that all these studies are based on the discretization method, the discrepancy became an open question. Our approach is completely independent and, therefore, resolves it.

We finish this discussion by directly comparing our method with the discretization codes. Let us consider the setup when we inject 70 MeV neutrinos with the amount $\rho_{\nu,\text{inj}}/\rho_{\nu,\text{tot}} = 5\%$. Fig.~\ref{fig:injection-70-MeV-5-perc} shows the evolution of $\delta \rho_{\nu}$ and neutrino spectra snapshot according to DSMC and the discretization codes from~\cite{Akita:2020szl,Boyarsky:2021yoh}, where for the latter we take the results shown in Fig. 7 from Appendix A. In the discretization codes, the electron mass effects are included.

We see a very similar behavior of the evolution predicted by DSMC and the discretization method from~\cite{Akita:2024ork}, both in terms of $\delta \rho_{\nu}$ and the spectrum. The tiny discrepancy may be explained by the fact that we have neglected the electron mass in the DSMC calculations. On the other hand, the discrepancy between DSMC and Ref.~\cite{Boyarsky:2021yoh} is somewhat larger. This is explained by the fact that the caption of Fig. 7 in Ref.~\cite{Boyarsky:2021yoh} wrongly mentions the setup other than the one actually used to make the plot. Unfortunately, the information about the true setup has been lost.

The running time of the DSMC simulation required to obtain Fig.~\ref{fig:injection-70-MeV-5-perc} is within a minute. In contrast, the discretization approach we used required $\simeq 10$ minutes. 

\subsubsection{Injection of 500 MeV neutrinos}

\begin{figure}
    \centering
    \includegraphics[width=0.45\textwidth]{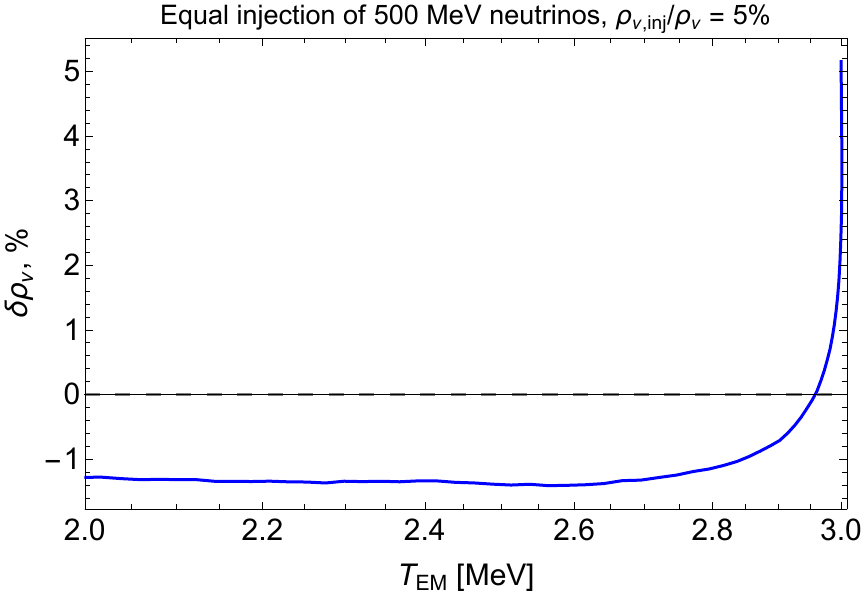}
    \caption{The same setup as in Fig.~\ref{fig:injection-20-MeV} but under an injection of 500 MeV neutrinos.}
    \label{fig:injection-500-MeV}
\end{figure}

Let us finalize this case study by considering the injection of 500 MeV neutrinos. The behavior of $\delta \rho_{\nu}$ is shown in Fig.~\ref{fig:injection-500-MeV}; it resembles the features shown in the case of the injection of 70 MeV neutrinos. 

The more important point is the performance of the DSMC setup. The running time required to simulate this setup was roughly the same as that for simulating 20 MeV and 70 MeV neutrinos. The 500 MeV case is already unrealistic to study using the discretization codes, as the running time would grow by a factor of $>(500/70)^{3}\simeq 400$. On the other hand, the running time of DSMC is roughly the same as for the 70 MeV scenario.

\subsection{Decays of long-lived SM particles}
\label{sec:application-mesons}

Let us now proceed with a more complicated case, when neutrinos are not injected directly in the decay chain but emerge via the evolution of heavy primary decay products $Y$, which may be muons or long-lived mesons such as $\pi^{\pm},K^{\pm},K_{L}$. 

In the primordial plasma, $Y$s experience a non-trivial evolution once being injected. The interactions they are involved in include kinetic energy loss, interactions with nucleons, annihilation with themselves, and decays, see~\cite{Akita:2024ta} for more details. The decay products generically involve neutrinos. This evolution influences their energy distribution among the neutrino and EM sectors.

\begin{figure}[h!]
    \centering
    \includegraphics[width=0.45\textwidth]{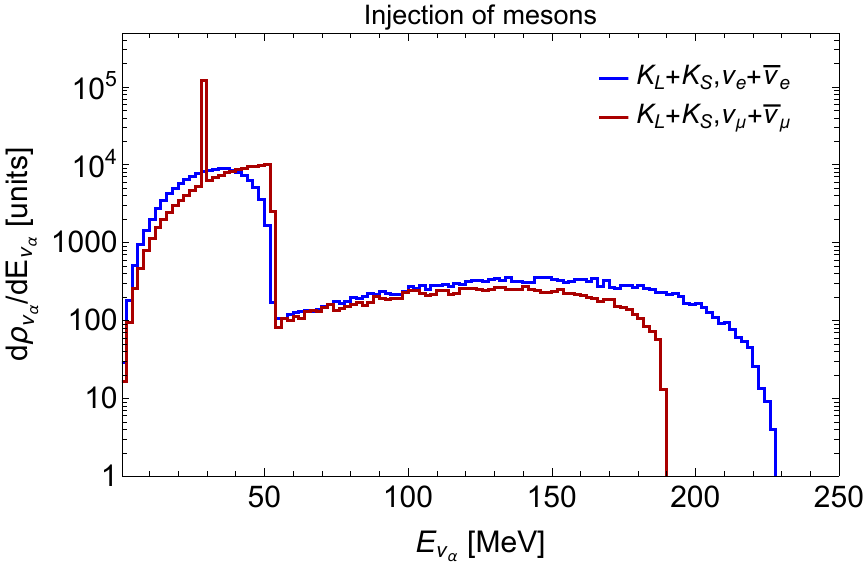} \\ \includegraphics[width=0.45\textwidth]{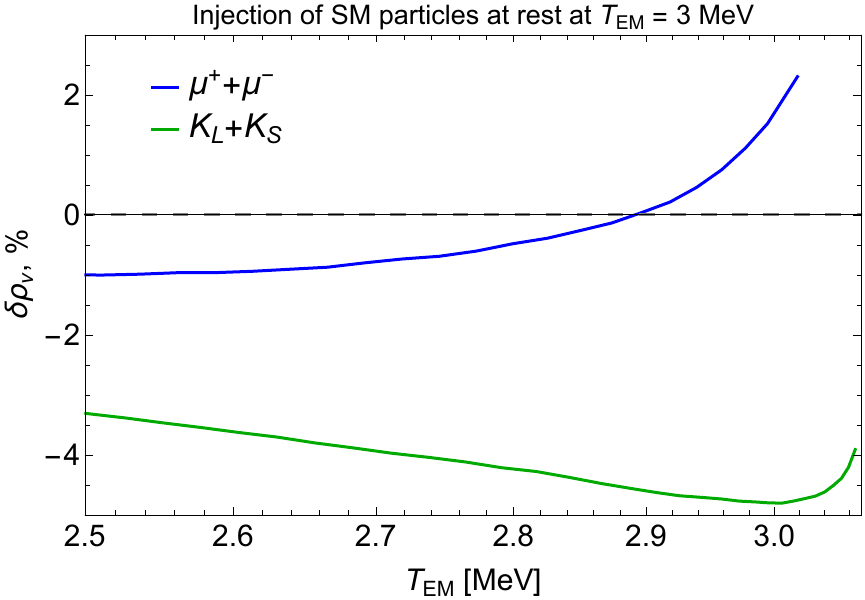}
    \caption{Impact of injection of heavy SM particles in the primordial plasma. \textit{Top panel}: the distribution of electron and muon neutrinos produced by decays of $K_{L}K_{S}$ pairs. When simulating their decay, we used the module of \texttt{SensCalc} tool~\cite{Ovchynnikov:2023cry}. For the chain of the decay products, we account for instant kinetic energy loss by charged particles. The continuous extension of the spectrum to $\simeq 200\text{ MeV}$ is caused by the direct decay of kaons into neutrinos. The increase at $E_{\nu} = 50\text{ MeV}$ follows from decays of secondary muons stopped in the plasma, whereas the sharp increase at $E_{\nu}\approx 34\text{ MeV}$ originates from decays of secondary pions. \textit{Bottom panel}: The evolution of the quantity $\delta \rho_{\nu}$ under the injection of $\mu^{+}\mu^{-}$ (the blue curve) and $K_{L}K_{S}$ (the green curve) in the primordial plasma at temperature $T_{\text{EM}} = 3\text{ MeV}$. The curves start at different temperatures $T_{\text{EM}}\neq 3\text{ MeV}$ because EM decays of these particles reheat the EM plasma.}
    \label{fig:neutrino-spectrum-KL}
\end{figure}

Our approach for simulating this redistribution is the following. We first inject these particles into the plasma and then decay them using Monte Carlo techniques. For the case of charged decay products, we transfer all the kinetic energy to the EM plasma and then decay them at rest. This is because the energy loss rate is much faster than any other relevant process in the MeV plasma. This simplified description follows the state-of-the-art studies~\cite{Fradette:2017sdd,Sabti:2020yrt}; the rest of the interactions discussed above will be added in the future. To simulate the phase space of the decay chain, we use \texttt{SensCalc}~\cite{Ovchynnikov:2023cry}, a tool calculating the event rate with the decaying LLPs at various laboratory experiments. It contains a module handling LLP decay chains and, in particular, the decays of different SM particles. We have modified it to incorporate the evolution of mesons and muons in the primordial plasma. In absolutely the same way, it may be used to simulate decays of the LLPs, with these mesons appearing among the final states.  

The neutrinos from $Y$s decays have a non-trivial spectrum. For instance, the neutrino distribution from decays of neutral kaons $K_{L}+K_{S}$ is shown in Fig.~\ref{fig:neutrino-spectrum-KL}. They have the following main decay modes:
\begin{align}
    K_{S}\to& 2\pi^{0}, 
    \quad K_{S}\to \pi^{+}\pi^{-}, \\ K_{L}\to& 3\pi^{0}, 
    \quad K_{L}\to \pi^{+}\pi^{-}\pi^{0}, 
    \quad K_{L}\to \pi^{\pm}l^{\mp}\nu_{l}
\end{align}
The neutral pions instantly decay into photons, just heating the EM 
plasma, whereas $\pi^{\pm},\mu^{\pm}$ particles lose kinetic energy before decaying:
\begin{equation}
    \pi^{+}\to \mu^{+}\nu_{\mu}, \quad \mu^{+}\to e^{+}\nu_{e}\bar{\nu}_{\mu}
\end{equation}
The spectrum of neutrinos from all these particles has the high-energy part with $E_{\nu}\gg T_{\text{EM}} = \mathcal{O}(\text{1 MeV})$, and we expect the same behavior of $\delta \rho_{\nu}$ as in the case of the injections of high-energy neutrinos. Clarifying this question is important since many past studies~\cite{Fradette:2017sdd,Fradette:2018hhl,Gelmini:2020ekg} treated these injections using the semi-analytic integrated Boltzmann approach (see, however, Refs.~\cite{Sabti:2020yrt,Rasmussen:2021kbf,Akita:2024ork}).

The evolution of $\delta \rho_{\nu}$ under the injection of $\mu^{+}\mu^{-}$ and $K_{L}K_{S}$ is shown in Fig.~\ref{fig:neutrino-spectrum-KL}. Let us start with the case of the muons. They inject $1/3$ of their energy into the EM plasma, with the rest going to the non-thermal neutrino population. Completely similar to the instant neutrino injection case, $\delta \rho_{\nu}$, being initially positive, instantly decreases below the $\Lambda$CDM value. This finding contradicts the studies~\cite{Fradette:2017sdd,Fradette:2018hhl}, which considered the scenario of decays of Higgs-like scalars into two muons and found that it increases $N_{\text{eff}}$ even in the regime of small scalar lifetimes $\mathcal{O}(0.1\text{ s})$. 

The $K_{L}K_{S}$ case is also interesting. Decaying, they put most of their energy into the EM plasma sector, so we start with a negative $\delta \rho_{\nu}$. However, the presence of very high-energy neutrinos with $E_{\nu}=100-200\text{ MeV}$ leads to a further slight drop of $\delta \rho_{\nu}$, and then it tries to approach the equilibrium.

\section{Conclusions}
\label{sec:conclusions}

Upcoming CMB observations will reach unprecedented precision, which may be used to discover or constrain new physics that was present in the primordial plasma at temperatures as large as a few MeV. To reach this goal, we have to understand the dynamics of the Early Universe in the presence of new physics. It requires solving the neutrino Boltzmann equation across a variety of scenarios, including long-lived relics, non-standard neutrino interactions, and lepton asymmetry in the neutrino sector. 

Current state-of-the-art methods are limited in scope and face computational challenges when neutrino evolution deviates significantly from the standard scenario. These limitations arise from the complex phase space of interactions, the presence of high-energy neutrinos, and the lack of analytic matrix elements -- features that are common in systems with new physics. Furthermore, the complexity of implementing these methods makes it difficult to extend them to include various new physics models, even within the range of applicability.

In this paper, we have presented an approach that is potentially free from all these limitations. It is based on the Direct Simulation Monte Carlo method to solve the Boltzmann equation, see Sec.~\ref{sec:DSMC-basics}. The traditional version of the DSMC approach is applied to rarefied gases and cannot be used to study the Early Universe. Fundamental modifications are required, such as including the Universe expansion, the hierarchy between weak and electromagnetic interaction rates, the Pauli principle, neutrino oscillations, and the presence of decaying particles. We have discussed these features and how to include them in the DSMC in Sec.~\ref{sec:DSMC-neutrinos}.

In Sec.~\ref{sec:current-implementation}, we have described our current proof-of-principle implementation of the DSMC approach for neutrinos that incorporates these modifications. We have validated it by conducting cross-checks within well-understood physics scenarios (see also Appendix~\ref{app:cross-checks}). In Sec.~\ref{sec:applications}, we have demonstrated the performance and flexibility of DSMC by applying its prototype to several toy scenarios that mimic real scenarios: the equilibration of the neutrinos and EM plasma initially having different temperatures (Sec.~\ref{sec:application-T}), injection of high-energy neutrinos (Sec.~\ref{sec:application-instant}), and injection of metastable particles including muons, pions, and kaons, which have a complicated decay chain including neutrinos (Sec.~\ref{sec:application-mesons}). 

Using these simple scenarios, we have found that the instant injection of high-energy neutrinos into a plasma with temperature $T_{\text{EM}}\gtrsim 1\text{ MeV}$ always decreases the neutrino-to-electromagnetic energy densities ratio, which leads to a negative change in $N_{\text{eff}}$ compared to the Standard cosmological scenario (see, in particular, Fig.~\ref{fig:injection-70-MeV-5-perc} and Fig.~\ref{fig:neutrino-spectrum-KL}). Being extended to the case of continuous neutrino injections in our companion paper~\cite{Ovchynnikov:2024xyd}, this finding resolves the previously existing discrepancy between different state-of-the-art approaches in predictions about the dynamics of $N_{\text{eff}}$ in the presence of high-energy neutrinos.

Our current neutrino DSMC code is rather proof-of-principle, limited by the efficiency of the implementation and some approximations. Once these problems are overcome, it will result in a powerful independent method of solving neutrino Boltzmann equations. We leave this for future work.

\section*{Acknowledgements}

We thank Stefan Stefanov for the in-depth review of the implementation of the proof-of-principle DSMC approach for neutrinos, and Fabio Peano, Luís Olivera e Silva, and Kyrylo Bondarenko for discussions at the early stages of this project. We also thank Kensuke Akita for helping with cross-checks of the DSMC approach, in particular, for adapting his approach to solve the neutrino Boltzmann equations for our needs. MO received support from the European Union's Horizon 2020 research and innovation program under the Marie Sklodowska-Curie grant agreement No. 860881-HIDDeN.

%For caption for Fig.\ref{fig:alp-production-probabilities} 

\newpage

\appendix

\newpage

\onecolumngrid 

\section{Scaling of the computational time of the discretization approach}
\label{app:discretization-scaling}

Let us first repeat the formula~\eqref{eq:discretization-scaling} for the scaling of the computational time of the discretization approach. We are interested in the domain from $t(T_{\text{ini}})$ to $t(T_{\text{fin}})$, when neutrinos with the energy $E_{\nu,\text{max}}$ are injected. The time is
\begin{equation}
    t_{\text{computation}} \propto N_{y}^{k+1}\times N_{\text{timestep}},
\end{equation}
where $N_{y}$ is the number of comoving momentum bins, and $N_{\text{timestep}}$ is the number of timesteps covering the time domain. $k$ is the minimal possible dimensionality of the collision integral under analytic reduction, remind Eq.~\eqref{eq:analytic-reduction}.

The power $N_{y}^{k+1}$ follows from the following considerations. Given $N_{y}$ bins, we have $N_{y}$ equations governing the evolution of the corresponding distribution modes. Next, each of the equations contains the collision integral, which, under the discretization, is represented as the product of $k$ connected summations over momentum modes:
\begin{equation}
    \mathcal{I}_{\text{coll},\nu_{\alpha}} = \int \prod_{i  =1}^{k} d\xi_{i} F(\{\xi\}) =\prod_{i = 1}^{k}\sum_{y_{i}}^{N_{y}}\Delta \xi_{i}F(\{\xi\})
\end{equation}

To obtain the scaling $t_{\text{computation}}(E_{\nu,\text{max}})$, we need to relate $N_{y}$ and $N_{\text{timestep}}$ to $E_{\nu,\text{max}}$.

\begin{itemize}
\item[--] In practice, for an arbitrary new physics model, the only robust choice of the binning is linear: $N_{y} = E_{\nu,\text{max}}/\Delta E$, with $\Delta E$ being the bin width (see a discussion below). $\Delta E$ must be kept constant to preserve the accuracy throughout the neutrino evolution. 
\item[--] Next, the timestep must be sufficiently small to resolve the neutrino equilibration rate. Consider the neutrino with the highest possible energy $E_{\nu,\text{max}}$. Its thermalization rate may be estimated using the Fermi theory as $\Gamma \sim n_{\nu}\cdot G_{F}^{2}\langle s\rangle \sim  G_{F}^{2}T^{4}E_{\nu,\text{max}}$, where $T$ is the plasma temperature. Therefore, as far as high-energy neutrinos are present in the plasma, the corresponding timestep scales as $\Delta t \sim \Gamma^{-1} \propto E_{\nu,\text{max}}^{-1}$. It means that to cover some fixed domain of time from $t(T_{\text{in}})$ and down to some moment $t(T_{\text{fin}})$, one would need $N_{\text{timestep}} \propto E_{\nu,\text{max}}$ timesteps. 
\end{itemize}
Therefore, the complexity grows as 
\begin{equation}
t_{\text{computation}} \propto E_{\nu,\text{max}}^{k+2}
\label{eq:complexity}
\end{equation}
Let us now briefly return to the choice of the binning. In principle, one may consider a different grid structure other than linear, e.g., logarithmic, or more exotic choices, see, e.g.~\cite{Gariazzo:2019gyi}. However, they would generically cause problems with energy conservation, accuracy, and stability for any beyond-the-standard scenario~\cite{Sabti:2020yrt,Akita:2024ork}. This is because, throughout the evolution, different comoving modes are populated mostly. The reason is that the spectrum of decaying LLPs is fixed in terms of physical momentum $p$, but varies in time if switching to the comoving momenta $y = a(t)\cdot p$. It is very unrealistic in practice to find an adjustment for a generic LLP: it would arbitrarily modify the dependence $a(t)$ and also decay into neutrinos with different energy distributions.

Another issue of the logarithmic grid is when there are two-body decays into neutrinos. Their energy distribution is just a $\delta$ function. In the comoving space, its argument moves towards different momenta $y$. Any binning other than linear would harm the accuracy when trying to resolve this peak.

\section{Details on the discretization approach}
\label{app:discretization-approach}
The discretization approach we use to compare with DSMC is discussed in Ref.~\cite{Akita:2024ork} (see also Ref.~\cite{Akita:2020szl}). It utilizes NLO QED corrections and includes three-flavor neutrino oscillations following Ref.~\cite{Sabti:2020yrt}.

Within the solver, we first introduce the following dimensionless variables: 
\begin{equation}
x=m_ea, \quad y=pa, \quad z=T_{\text{EM}}a,
\end{equation}
normalizing $z\rightarrow 1\ (a\rightarrow 1/T_\text{EM})$ at the high-temperature limit. The quantities $x,y,z$ characterize time, momentum, and photon temperature, respectively. Then, we discretize the comoving momentum. The discretization is linear: for the neutrino (electromagnetic) momentum grid $y_i$, we use 200 (80) grid points with $y_{\rm min}=0.01\ (0.01)$ and $y_{\rm max}={\rm max}[a_{\rm stop}m_X/2, 40]\ (40)$, where $a_{\text{stop}}$ is the estimate of the final scale factor, and $m_{X}$ is the LLP's mass.

The Boltzmann solver is written in \texttt{Python} with \texttt{scipy}, \texttt{numpy} and \texttt{numba} libraries as in~\cite{Froustey:2020mcq}. To solve the ordinary differential equations on the discretized neutrino modes, we use the \texttt{LSODA} method in \texttt{solve\_ivp} distributed in \texttt{scipy}. By considering the setups with continuous injections of neutrinos of energy $E_{\nu,\text{max}}$, we have recovered the approximate scaling~\eqref{eq:discretization-scaling}.\footnote{The scaling is slightly worse because of the need for computing the jacobian of the system of ODEs computed within \texttt{LSODA}.} 

\section{Convergence of the DSMC algorithm}
\label{app:convergence-tests}

\begin{figure}[h!]
    \centering
    \includegraphics[width=0.5\linewidth]{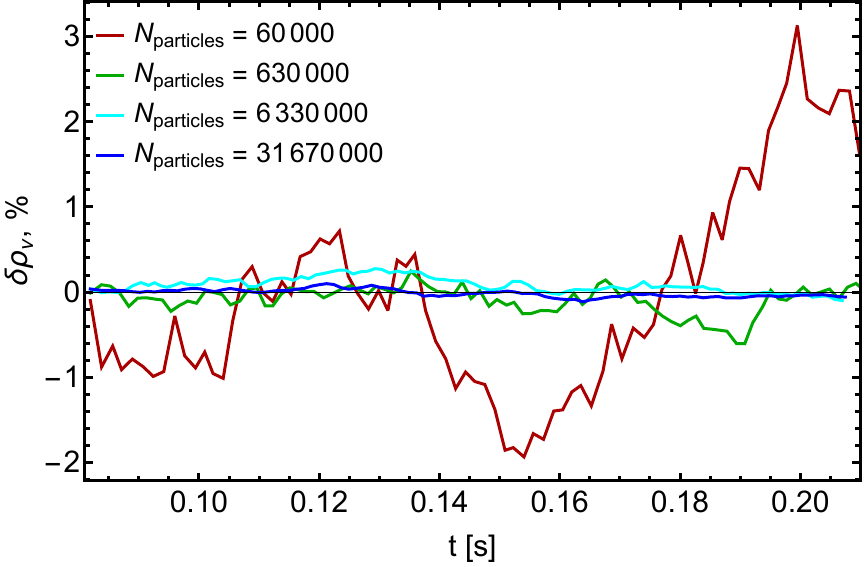}~\includegraphics[width=0.5\linewidth]{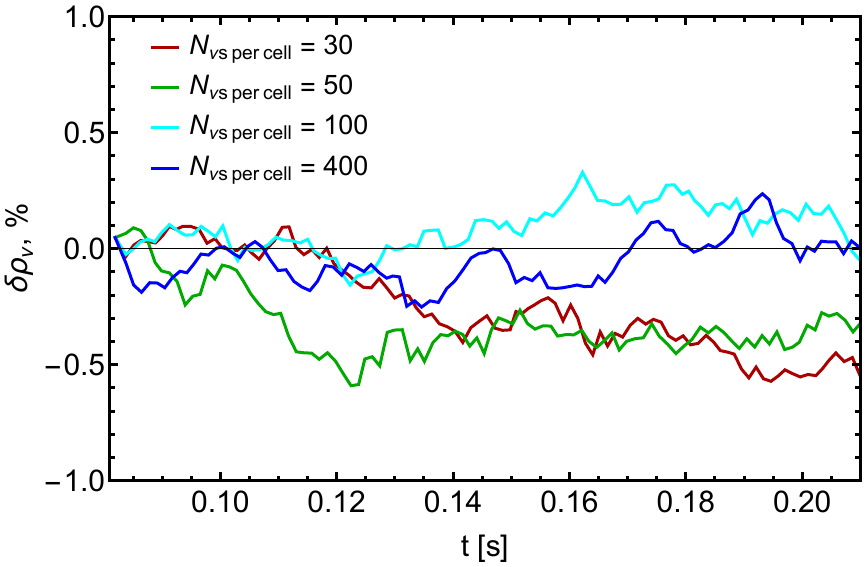}
    \caption{The temporal evolution of the quantity $\delta \rho_{\nu}$ when varying numbers of neutrinos per cell $N_{\text{cell},\nu}$ and particles in the system $N$. \textit{Left panel}: fixing $N_{\text{cell},\nu} = 400$ and varying $N$. \textit{Right panel}: fixing $N = 3\cdot 10^{6}$ and varying $N_{\text{cell},\nu}$.}
    \label{fig:convergence}
\end{figure}

In order to have robust DSMC simulations, we have to use large enough numbers of simulated particles $N$ and particles per cell, $N_{\text{cell}}$ (here and below, we assume that the computational particles match the physical particles). The former is needed to overcome the Monte-Carlo noise -- random fluctuations of macroscopic observables around their expected values. The latter is crucial because the simulation in each cell involves the calculation of temperatures of the EM plasma and effective neutrino temperatures. If the number of particles of the given type (EM particles, or neutrinos $\nu_{\alpha}$) per cell is too small, the temperature may have large statistical fluctuations, adding noise on top.

In Fig.~\ref{fig:convergence}, we test the system's behavior with neutrinos and EM particles that are initially in perfect equilibrium. We consider two setups: 
\begin{itemize}
\item[1.] The one with $N_{\text{cell}} = 400$ and $N$ varying from $N = 6\cdot 10^{4}$ to $N = 3\cdot 10^{7}$.
\item[2.] Another one with $N = 5\cdot 10^{3}$ and the number of neutrinos per cell $N_{\nu,\text{per cell}}$ ranging from 30 to 400. 
\end{itemize}
Our goal is to define the setup with the noise at the sub-percent level, which keeps the system in the dynamic equilibrium.

For the first setup, we observe the random fluctuations of the quantity $\delta \rho_{\nu}$, defined by Eq.~\eqref{eq:delta-rho}, within 2-3\% around zero. The simulation with $N = 3\cdot 10^{7}$ has fluctuations at the level $\mathcal{O}(0.1\%)$, which roughly corresponds to the scaling of the fluctuations as $1/\sqrt{N}$. It is well enough for our purposes.

We can reach large $N$ either considering a single DSMC simulation with this $N$, or, equivalently, averaging over $n$ simulations with the number of particles $N/n$. This flexibility allows using DSMC even on laptops without large RAM and simultaneously accumulating large $N$.

For the second setup, considering small values of $N_{\text{cell}}$, we not only gain additional fluctuations but constantly drive $\delta \rho_{\nu}$ towards negative values. Its origin is rounding the number of the EM particles after updating the local EM cell temperature. If this number is tiny (which is the case when $N_{\text{cell}}$ is small), rounding causes a statistically significant effect. The problem gradually disappears once $N_{\text{cell}}\gtrsim 100$. We will consider $N_{\text{cell}} = 400$, because it provides a balance between the performance of the code and the quality of statistical sampling.

Having defined the stable setup in the perfect equilibrium case, we can now pose the question of whether it is stable if injecting high-energy neutrinos. Let us assume that these neutrinos have the energy $E_{\nu,\text{inj}}$ carry the fraction $\Delta$ of the energy density of the thermal plasma; in practice, $\Delta$ is fixed by the initial energy density of the decaying LLP. Fixing the number of thermal particles $N$ and estimating their mean energy as $\langle E\rangle = 3.15 \cdot T_{\text{EM}}$, we can derive the number of injected neutrinos $N_{\text{injected}}$:
\begin{equation}
    N_{\text{injected}} = \frac{3.15\cdot \Delta \cdot N\cdot T_{\text{EM}}}{E_{\nu,\text{inj}}}
\end{equation}
For the given $N,T_{\text{EM}},E_{\nu,\text{inj}}$, $N_{\text{injected}}$ has to be large enough. Otherwise, fluctuations are possible: if the number of injected neutrinos is too small, they may all annihilate into the EM plasma particles during the first interaction.

In practice, we have found that for $N\gtrsim 10^{6}$, the number $N_{\text{inj}}$ for which the fluctuations are manageable is $N_{\text{inj}}\gtrsim 100$. For instance, for $T_{\text{EM}} = 3\text{ MeV}$ and $\Delta = 0.05$, we satisfy this requirement as far as $E_{\nu,\text{inj}} \lesssim 10\text{ GeV}$.

Even if we cross this extreme limit, we may overcome fluctuations if increasing $N$ \textit{linearly} with $E_{\nu,\text{inj}}$. In this case, the scaling of the computational time of the DSMC algorithm would be $t_{\text{computation}}\propto E_{\nu,\text{inj}}^{2}$, where one power comes from the number of timesteps and another from increasing $N$. This is still way better than the scaling~\eqref{eq:complexity} of the discretization approach, which is \textit{at least} $\propto E^{4}_{\nu,\text{inj}}$.

\section{Cross-checks}
\label{app:cross-checks}

\subsection{Approaching thermal equilibrium} 

To test whether the DSMC simulation brings the system of neutrinos and EM particles to the dynamical equilibrium defined by Eqs.~\eqref{eq:rho-distribution-thermal} and~\eqref{eq:rho-ratio-thermal}, we will use the following setup: 
\begin{itemize}
\item[--] The Universe contains neutrinos and anti-neutrinos of all flavors together with electrons, positrons, and photons.
\item[--] The expansion of the Universe is absent. Therefore, the total energy density of the system is constant, the particles' momenta do not experience redshift and are subject only to their interactions. Such an assumption is needed to allow neutrinos to thermalize fully, making interpreting the results more transparent.
\item[--] The initial distribution function of neutrinos consists of two components that are the same for all flavors: 
\begin{itemize}
    \item[1.] The equilibrium component, which has Fermi-Dirac distribution with the temperature $T_{\nu}^{\text{ini}} = 3 \text{ MeV}$.
    \item[2.] The non-equilibrium component - neutrinos with an arbitrary energy distribution, with the energy density constituting some fraction $\ll 1$ of the equilibrium energy density.
\end{itemize}
\end{itemize}
The first sub-scenario we consider is where there are no non-equilibrium neutrinos, so the system is initially in the equilibrium state. If at least one component of the DSMC simulation is implemented incorrectly, the system will escape the equilibrium, tending to the false ground state. A prominent example is when the cross-sections are taken to be velocity-independent; then, the distribution of the system tends to the fake-equilibrium spectrum
$d\rho_{\nu}/dE_{\nu} \sim E_{\nu}^{2}\times f_{\text{FD}}$ instead of the correct $E^{3}_{\nu}f_{\text{FD}}$ (for relativistic particles with Boltzmann statistics, such an issue has been encountered and explained in~\cite{peano2009statistical}). Another issue may be if the maximal interaction weight $\omega_{\text{max}}$ in the acceptance criterion of the pair's interaction~\eqref{eq:interaction-acceptance} is not actually the maximal one. Then, the system falls into the state with $\delta \rho_{\nu} < 0$.

\begin{figure}
    \centering
    \includegraphics[width=0.45\textwidth]{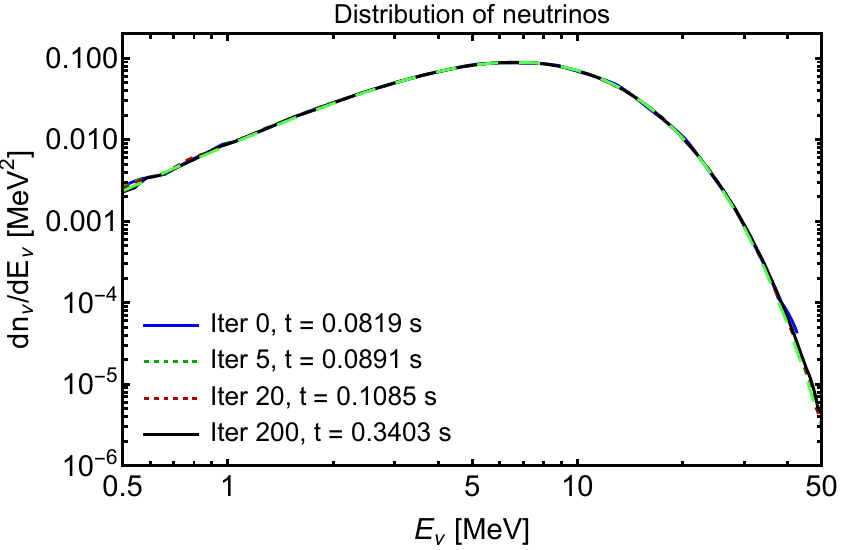}
    \caption{The evolution of the neutrino distribution $dn_{\nu}/dE_{\nu}$ averaged over all flavors under the assumption of fully equilibrium initial conditions~\eqref{eq:rho-distribution-thermal} and~\eqref{eq:rho-ratio-thermal}. The ``Iter \#'' curves correspond to the number of the iteration. No significant changes are developed throughout the simulation. The minor changes are related to the quality of the sampler of the kinematics of the electrons via the Fermi-Dirac distribution. The dashed green line shows the analytic Fermi-Dirac distribution with the temperature equal to the temperature of the electromagnetic plasma $T_{\text{EM}}$.}
    \label{fig:Equilibrium_consistency}
\end{figure}

Our DSMC implementation passes this test, see Fig. \ref{fig:Equilibrium_consistency}.

\begin{figure*}[t!]
    \centering
    \includegraphics[width=0.5\linewidth]{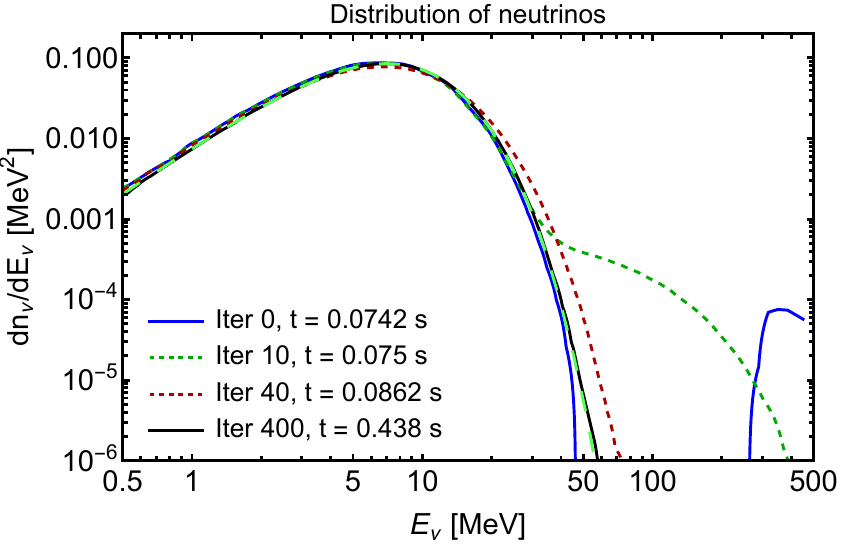}~\includegraphics[width=0.5\linewidth]{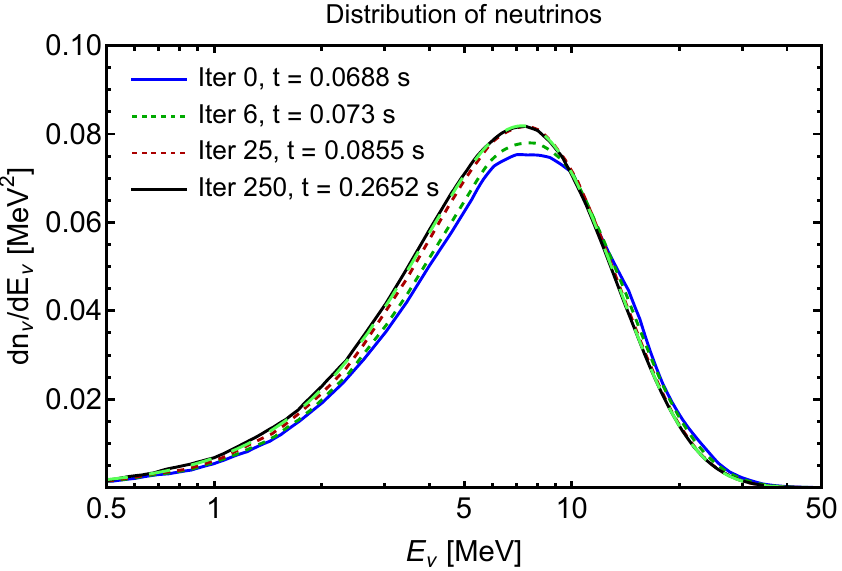}
    \caption{Evolution of the neutrino distribution function $dn_{\nu}/dE_{\nu}$ averaged over all flavors under different initial setups, showing how DSMC drives it towards thermal equilibrium with the EM sector. \textit{Left panel}: with equilibrium neutrinos and EM plasma at temperature $T_{\text{EM}} = 3\text{ MeV}$ and non-equilibrium neutrinos with energies randomly distributed in the range $300 \text{ MeV} < E_{\nu} < 450 \text{ MeV}$. Their total energy density is related to the total energy of the equilibrium part as $\rho_{\nu_\alpha}^{\text{non-eq}}/\rho_{\nu}^{\text{eq}} = 0.15$. The non-equilibrium part of the spectra rapidly loses its energy in the first steps of simulation, leading to the distortions of the spectra at high energies, which are eventually equilibrated. The plot shows the snapshots of the binned neutrino distribution function as obtained at different iterations of the DSMC simulation. The iteration 0 corresponds to the initial setup, while the iteration 400 is the final state. For comparison, the long-dashed green line shows the Fermi-Dirac distribution $dn_{\nu}/dE_{\nu} = E_{\nu}^{2}f_{\text{FD}}(E_{\nu},T_{\text{EM,final}})$, being the thermal equilibrium of neutrinos with the EM plasma with the final temperature $T_{\text{EM,final}} \approx 3.15\text{ MeV}$. \textit{Right panel}: with equilibrium neutrinos having temperature $T_{\nu_{\alpha}} = 3.5\text{ MeV}$ and EM plasma at temperature $T_{\text{EM}} = 3\text{ MeV}$. The meaning of the lines is the same, while the number of iterations is 250.}
\label{fig:Equilibrium_test}
\end{figure*}

Next, we consider two non-trivial initial conditions: different temperatures of neutrinos and EM particles, and the addition of non-equilibrium neutrinos. The relaxation of the neutrino distribution to the equilibrium one for such scenarios is shown in Fig.~\ref{fig:Equilibrium_test}. Its results are in perfect agreement with the theoretical expectations.

\subsection{Energy transition rates} 

In this scenario, we will reproduce the semi-analytical result of \cite{Escudero:2018mvt,EscuderoAbenza:2020cmq}, where the evolution of neutrinos in the Early Universe was studied under an assumption that every moment of time, the shape of their energy distribution is thermal. The energy transition rates were calculated analytically in terms of the temperatures of neutrinos and EM plasma $T_{\nu_{\alpha}},T$. The Boltzmann equations are reduced to the simple system of differential equations on $T_{\nu_{\alpha}},T$. For our simulation, the following setup will be used:

\begin{figure}[h!]
    \centering
    \includegraphics[width=0.45\linewidth]{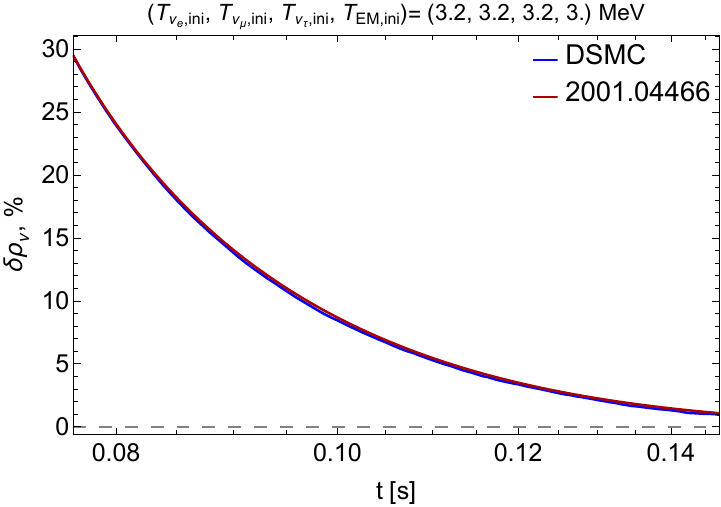}~\includegraphics[width=0.45\linewidth]{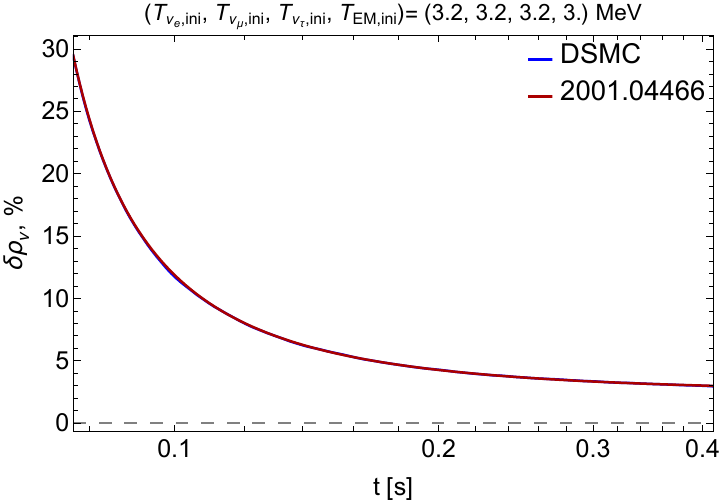}
    \caption{The evolution of the ratio of the neutrino energy density to the EM energy density in DSMC simulation compared to the theoretical prediction from~\cite{EscuderoAbenza:2020cmq}, under an assumption that the shape of the neutrino distribution function is always thermal at each step of the simulation. The initial conditions for the setup are $T_{\nu_i} = 3.2$ MeV for every flavor and the temperature of the EM plasma is $T_{\text{EM}} = 3$ MeV. \textit{Left panel}: not including the expansion of the Universe. Due to the absence of expansion, the ratio approaches to the exact SM value. \textit{Right panel}: expansion included.}
    \label{fig:Rates_test}
\end{figure}

\begin{itemize} 
\item[--] The Universe's content is neutrinos and anti-neutrinos of all flavors together with electrons, positrons, and photons. 
\item[--] The simulation is altered such that neutrino distributions always have the shape~\eqref{eq:rho-distribution-thermal} at each simulation step. Basically, we treat neutrinos in exactly the same way as the EM particles in the full DSMC simulation. 
\item[--] The expansion of the Universe is not included to concentrate on the energy exchange rates.
\item[--] As in the whole study, the electron mass is set to zero.
\end{itemize}
The example of the resulting evolution of the energy density of the neutrino plasma is presented in Fig.\ref{fig:Rates_test}, where the almost perfect correspondence between theoretical predictions and simulation can be seen. Such reproduction of the energy evolution behavior confirms that averaged energy transition rates are computed correctly.

\subsection{Expansion and decoupling}

In the third cross-check, we will follow the previous setup, but with the expansion of the Universe included. Due to initial difference between temperatures of neutrino and EM plasma, we expect some remaining inequality between them, since the start of the simulation occurs close to the temperature of the neutrino decoupling. In similar terms, we present the example of such comparison in Fig.~\ref{fig:Rates_test}.

\bibliography{main.bib}

\end{document}